\documentclass[iop, twocolumn, appendixfloats, twocolappendix, numberedappendix]{emulateapj}
\usepackage{apjfonts}
\usepackage{natbib}
\usepackage{amsfonts}
\usepackage{amsmath}
\usepackage[usenames]{color}
\usepackage{verbatim}
\usepackage{graphicx}
\usepackage{subfigure}

\newcommand\msun{{M}_\odot}
\newcommand\kms{km\;s$^{-1}$}
\renewcommand\aa{\AA}

\shorttitle{}
\shortauthors{}

\begin{document}
\title{The Assembly Histories of Quiescent Galaxies Since {\lowercase{\it z}} = 0.7 from Absorption Line Spectroscopy}
\author{Jieun Choi\altaffilmark{1}}
\author{Charlie Conroy\altaffilmark{1}}
\author{John Moustakas\altaffilmark{2}}
\author{Genevieve J. Graves\altaffilmark{3}}
\author{Bradford P. Holden\altaffilmark{4}}
\author{Mark Brodwin\altaffilmark{5}}
\author{Michael J. I. Brown\altaffilmark{6}}
\author{Pieter G. van Dokkum\altaffilmark{7}}

\altaffiltext{1}{Department of Astronomy \& Astrophysics, University of California, Santa Cruz, CA 95064, USA}
\altaffiltext{2}{Department of Physics and Astronomy, Siena College, Loudonville, NY 12110, USA}
\altaffiltext{3}{Department of Astrophysical Sciences, Princeton University, Princeton, NJ 08544, USA}
\altaffiltext{4}{UCO/Lick Observatories, University of California, Santa Cruz, CA 95064, USA}
\altaffiltext{5}{Department of Physics and Astronomy, University of Missouri, Kansas City, MO 64110, USA}
\altaffiltext{6}{School of Physics, Monash University, Clayton, Vic 3800, Australia}
\altaffiltext{7}{Department of Astrophysical Sciences, Yale University, New Haven, CT 06520, USA}

\begin{abstract}
We present results from modeling the optical spectra of a large sample of quiescent galaxies between $0.1<z<0.7$ from the Sloan Digital Sky Survey (SDSS) and the AGN and Galaxy Evolution Survey (AGES). We examine how the stellar ages and abundance patterns of galaxies evolve over time as a function of stellar mass from $10^{9.6}\text{--}10^{11.8}~\msun$. Galaxy spectra are stacked in bins of mass and redshift, and modeled over a wavelength range from 4000~\aa{} to 5500~\aa{}. Full spectrum stellar population synthesis modeling provides estimates of the age and the abundances of the elements Fe, Mg, C, N, and Ca. We find negligible evolution in elemental abundances at fixed stellar mass over roughly 7~Gyr of cosmic time. In addition, the increase in stellar ages with time for massive galaxies is consistent with passive evolution since $z=0.7$. Taken together, these results favor a scenario in which the inner $\sim0.3\text{--}3\;R_{\rm e}$ of massive quiescent galaxies have been passively evolving over the last half of cosmic time. Interestingly, the derived stellar ages are considerably younger than the age of the universe at all epochs, consistent with an \emph{equivalent} single-burst star formation epoch of $z\lesssim1.5$. These young stellar population ages coupled with the existence of massive quiescent galaxies at $z>1$ indicate the inhomogeneous nature of the $z\lesssim0.7$ quiescent population. The data also permit the addition of newly-quenched galaxies at masses below $\sim10^{10.5}~\msun$ at $z<0.7$. Additionally, we analyze very deep Keck DEIMOS spectra of the two brightest quiescent galaxies in a cluster at $z = 0.83$. There is tentative evidence that these galaxies are older than their counterparts in low-density environments. In the Appendix, we demonstrate that our full spectrum modeling technique allows for accurate and reliable modeling of galaxy spectra to low S/N ($\sim20$~\aa{}$^{-1}$) and/or low spectral resolution ($R\sim500$). 
\end{abstract}


\maketitle
\section{INTRODUCTION}
Considerable progress has been made over the last several decades toward understanding the formation and assembly histories of galaxies. On the observational front, two complementary techniques are widely used. Lookback studies aim to study the evolution of galaxies in a statistical manner by comparing snapshots of the galaxy population at different cosmic epochs. This technique has been powered by major spectroscopic surveys such as the Sloan Digital Sky Survey \citep[SDSS;][]{York2000}, the 2dF Galaxy Redshift Survey \citep[2dFGRS;][]{Colless2001}, the VIMOS VLT Deep Survey \citep[VVDS;][]{LeFevre2004, LeFevre2005}, the PRIsm MUlti-object Survey \citep[PRIMUS;][]{Coil2011}, the Deep Extragalactic Evolutionary Probe 2 (DEEP2) survey \citep{Newman2013}, and the AGN and Galaxy Evolution Survey \citep[AGES;][]{Kochanek2012}, among many others. In addition, targeted surveys out to $z\sim1$ have provided many insights into the assembly histories of galaxies \citep[e.g.,][]{vanDokkum1996, Rusin2005, Treu2005, Holden2010, Jorgensen2013}. In contrast, in the archaeological approach, one infers past evolution through detailed studies of $z\sim0$ galaxies and their stellar populations. This method of extrapolating back in time is enabled by high quality data of nearby galaxies.

One of the most basic probes of galaxy formation and evolution enabled by large-scale redshift surveys is the time evolution of galaxy luminosity and stellar mass functions. At $z\gtrsim2$, star-forming galaxies dominate quiescent galaxies in number at all stellar masses, but the mass density of quiescent galaxies has increased by almost an order of magnitude between $z\sim2$ and today \citep[e.g.,][]{Bell2004, Blanton2006, Brown2007, Faber2007, Cirasuolo2007, Arnouts2007, Cappellari2009, Ilbert2010, Whitaker2010, vandeSande2011, Brammer2011, DominguezSanchez2011}. While the quiescent population has been on a global rise, there is strong evidence that this process is mass-dependent. Between $z\sim2$ and $z\sim1$, the space density of massive galaxies has been observed to increase rapidly \citep{Arnouts2007, Cirasuolo2007, Ilbert2010, Nicol2011, Brammer2011}. By $z\sim1$, massive galaxies are mostly assembled, and they appear to passively evolve to $z\sim0$ \citep[e.g.,][]{Bundy2006, Renzini2006, Cirasuolo2007, Vergani2008, Marchesini2009, Banerji2010, Moustakas2013, Muzzin2013}. Thus the rapid evolution in the mass and luminosity functions of quiescent galaxies at late times has mostly been attributed to the rise in low- and intermediate-mass quiescent galaxies, although the details are still under active debate \citep[e.g.,][]{Cimatti2006, Scarlata2007, Brown2007, Stewart2009, Ilbert2010, Robaina2010, Whitaker2010, ElicheMoral2010, Pozzetti2010, Brammer2011, Skelton2012, Moustakas2013}. However, mass and luminosity functions by necessity depend on the ability to accurately measure global photometry of extended sources. In particular, accurate photometry accounting for diffuse light at large radius is notoriously difficult even at $z=0$ \citep{Bernardi2013}, and it only becomes more challenging at higher redshifts due to the $(1+z)^{4}$ decrease in surface brightness. 

There has been a tremendous effort in the past decade to study the outskirts of massive quiescent galaxies as a function of cosmic time. Both size evolution studies using deep imaging and dynamical studies have shown that the sizes of massive galaxies have increased by a factor of $2\text{--}4$ from $z\sim2$ to the present \citep{Daddi2005, Trujillo2006, vanDokkum2008, vanderWel2008, Cimatti2008, Bezanson2009, Damjanov2009, Williams2010, Cassata2010, vanDokkum2010, LopezSanjuan2012, McLure2013, Belli2013}. The inner regions ($r\lesssim5$ kpc) of massive galaxies have apparently undergone very little mass growth, but mass out to $\sim75$ kpc has increased by a factor of four since $z\sim2$ \citep[e.g.,][]{vanDokkum2010}. The main channel for size and mass growth is likely dominated by minor mergers, though in-situ star formation and major mergers are also thought to play a role. A simple scaling argument based on the virial theorem explains the dramatic size growth during minor mergers, where the radius increases quadratically with mass instead of linearly as in major mergers \citep{Naab2009}. Observations showing that the oldest and most massive galaxies at high redshift are smaller by a factor of $\sim5$ compared to low-redshift galaxies of comparable mass \citep[e.g.,][]{Daddi2005, Onodera2012} further corroborate the ``inside-out growth'' of massive galaxies. Simulations support this notion that galaxy formation occurs in two phases: compact cores are thought to form rapidly at $z\gtrsim2$ from star formation triggered by infalling cold gas, followed by a slower growth in both mass and size over a longer period of time through the accretion of satellites \citep[e.g.,][]{Naab2007, Keres2009, Naab2009, Hopkins2009a, Dekel2009, Lackner2012, Hilz2013}. This is still a controversial field, however, with some groups proposing other modes of size growth, e.g., baryonic mass loss leading to the ``puffing up'' of galaxies \citep[e.g.,][]{Fan2010}. Others invoke progenitor bias, arguing that the new galaxies entering the quiescent population are inherently larger in size \citep[e.g.,][]{Carollo2013}, and some groups even find evidence for a lack of size evolution \citep[e.g.,][]{Mancini2010, Stott2011}.

Stellar population analysis offers yet another way of probing the evolution of galaxies (see \citealt{Walcher2011}, \citealt{Conroy2013} for recent reviews). The most popular technique for analyzing properties of old stellar populations in quiescent galaxies, such as age, metallicity, and abundance patterns, is to measure and model several key absorption features using the Lick/IDS index system \citep{Burstein1984, Worthey1994}. Based on this approach, stellar populations in massive quiescent galaxies are found to be old and enhanced in $\alpha$ elements compared to the Milky Way disk stars \citep[e.g.,][]{Worthey1994}. Moreover, numerous independent groups have found strong positive correlations with velocity dispersion for age, total metallicity, and the ratio of $\alpha$ elements to Fe---almost ubiquitously represented by [Mg/Fe] in the literature---but almost no trend for [Fe/H] \citep{Trager1998, Thomas2005, Graves2007, Schiavon2007, Smith2009, Zhu2010, Johansson2012, Conroy2014, Worthey2014}. The total metallicity is an important diagnostic for galaxy formation and evolution because it is sensitive to the depth of the potential well in which their stellar populations formed \citep[e.g., supernovae-driven winds can efficiently remove metals from shallow potential wells;][]{Larson1974}. On the other hand, [$\alpha$/Fe] is sensitive to the timescale of star formation: massive stars expel $\alpha$ elements into their interstellar environments on million-year timescales, while Type Ia supernovae enrich the star-forming gas with Fe on billion-year timescales \citep[e.g.,][]{Tinsley1979}. By measuring the relative abundances of $\alpha$ elements and Fe in stellar populations, the time and the rate at which stars formed in their host galaxies can be inferred \citep[e.g.,][]{Thomas1999}.

More recently, several groups have begun modeling the full optical spectrum of galaxies. First used to measure star formation histories and metallicities \citep{Heavens2000, CidFernandes2005, Ocvirk2006, Tojeiro2009}, this technique was further developed by \cite{Walcher2009} and \cite{Conroy2012} to include variable elemental abundances. The present work can be viewed as the high-redshift extension of \cite{Conroy2014}, which focused on full spectrum modeling of high-quality SDSS spectra of $z\sim0$ quiescent galaxies to derive their ages and detailed abundance patterns. In this work, we use a hybrid approach of combining lookback and archaeological studies to carry out detailed stellar population analysis at progressively higher redshifts. For the first time, we derive accurate ages and abundance measurements for a large mass-complete sample of galaxies from $z\sim0.1$ to $z\sim0.7$, and examine how their stellar population properties evolve over time as a function of stellar mass. As we demonstrate, powerful constraints can be placed on the assembly histories of galaxies from the time evolution of their stellar population properties.

The paper is organized as follows: Section~\ref{section:data} gives an overview of the data sets and the sample selection process, and Section~\ref{section:model} presents background information on our models and the full spectrum fitting technique. The main science results are introduced in Section~\ref{section:results}, and a discussion and summary are provided in Sections~\ref{section:discussion} and \ref{section:summary}. We discuss the results of various systematic tests to explore the robustness of our analysis in the Appendix. All wavelengths in this paper are quoted in vacuum. Where necessary, we assume a Chabrier IMF \citep{Chabrier2003} for the stellar mass range \mbox{0.1--100~$\msun$} unless noted otherwise, and $\Lambda$CDM cosmology, with \mbox{$H_0$ = 70 \kms{} Mpc$^{-1}$}, $\Omega_{\rm m} = 0.3$, and $\Omega_{\rm \Lambda} = 0.7$.

\begin{figure*}[]
\centering
\includegraphics[width=1.7\columnwidth]{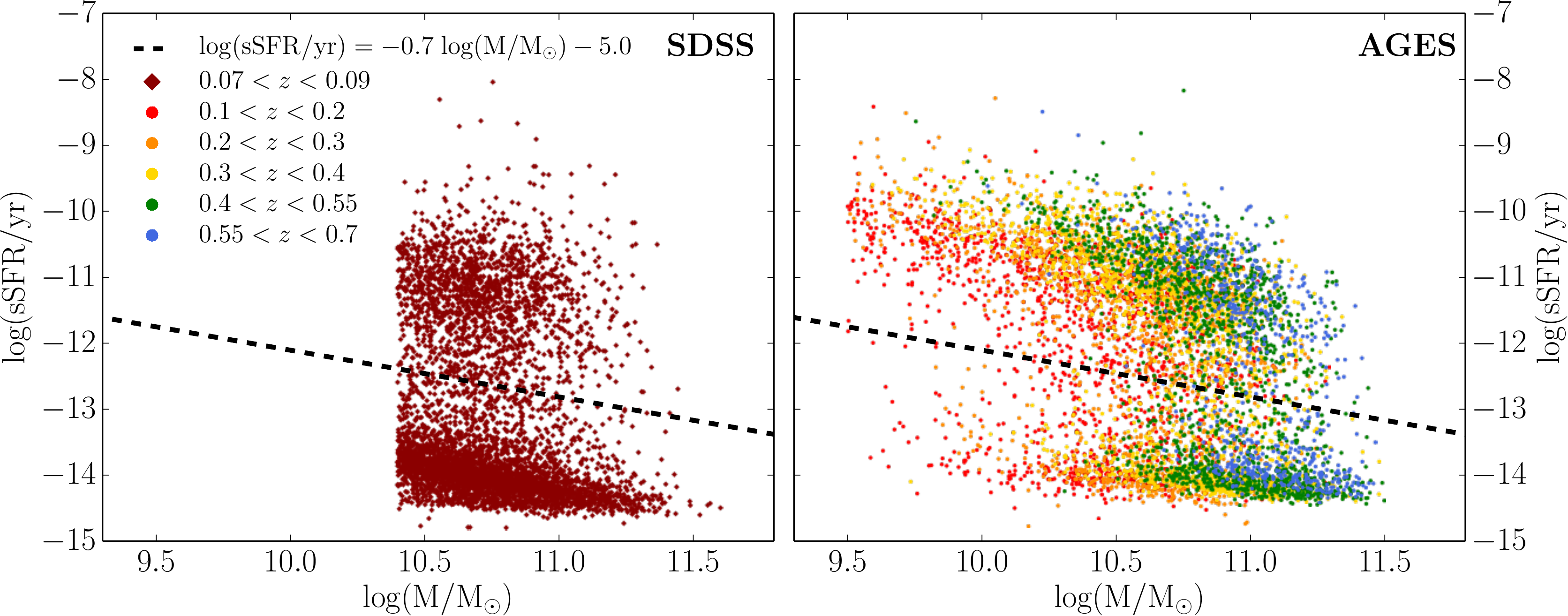}
\caption{{\it Left panel}: specific star formation rate (sSFR) as a function of stellar mass for a subset of the SDSS sample. A random subset is shown for display purposes. Both the quiescent cloud and the star-forming main sequence can be seen. We locate the minimum of the bimodal distribution in three mass slices and perform a linear least-squares fit to those points. The resulting quiescence threshold is shown as a dashed line. We discard all galaxies above this quiescence cut to obtain a sample of quiescent galaxies. {\it Right panel}: sSFR as a function of stellar mass for the entire AGES sample. The black dashed line corresponds to the quiescence cut determined from the SDSS sample. Only those that fall below the sSFR threshold are included in our quiescent sample. We use a quiescence cut that is constant with redshift because there is no evidence for evolution in the location of the green valley.}
\label{fig:sSFR_vs_mass}
\end{figure*}

\section{DATA}
\label{section:data}
In this work we primarily utilize two large spectroscopic surveys to select samples of quiescent galaxies. At intermediate redshift we rely on the AGN and Galaxy Evolution Survey \citep[AGES;][]{Kochanek2012}, while at low redshift we use a large sample of galaxies selected from the Sloan Digital Sky Survey \citep[SDSS;][]{York2000}. In this section we provide brief overviews of the surveys, data, and sample selection process.

\subsection{SDSS}
\label{section:sdss_sample}
Our low-redshift sample of galaxies is selected from the SDSS Data Release 7 \citep[DR7;][]{Abazajian2009}. This dataset includes $ugriz$ photometry for 357 million objects over 11,663~deg$^2$, and optical spectroscopy for roughly $930,000$ galaxies over 9380 deg$^2$. We choose $427,536$ galaxies observed as part of the main galaxy survey \citep{Strauss2002} with $14.5 < r < 17.6$ and $0.05 < z <0.2$ from the SDSS/DR7 New York University Value Added Galaxy Catalog \citep[NYU/VAGC;][]{Blanton2005}.\footnote{{\url{http://sdss.physics.nyu.edu/vagc}}}

Following \cite{Moustakas2013}, broadband photometry from the ultraviolet (UV) to the rest-frame near-infrared is assembled for this sample. Specifically, we obtain near- and far-UV photometry from the \emph{Galaxy Evolution Explorer} \citep[\emph{GALEX};][]{Martin2005} \emph{GALEX} Release 6 (GR6), optical $ugriz$ photometry from the SDSS, and \emph{Wide-field Infrared Survey Explorer} \citep[\emph{WISE};][]{Wright2010} [3.4] and [4.6] photometry from the \emph{WISE} All-Sky Data Release. We make every effort to ensure that our photometry captures as much of the total light of the galaxy in each band as possible; in particular, the SDSS {\tt model} magnitudes are used to measure the optical colors of each galaxy, scaled by the $r$-band {\tt cmodel} magnitude, which provides the most reliable estimate of the integrated (total) galaxy flux irrespective of galaxy type \citep{Bernardi2010, Blanton2011}. Finally, the stellar velocity dispersions measured by the SDSS pipeline and distributed as part of the NYU/VAGC are adopted in this work.

We use {\tt iSEDfit} \citep{Moustakas2013}, a Bayesian SED modeling code, to derive the stellar masses and star formation rates for the galaxies in our sample. The Flexible Stellar Population Synthesis models \citep[FSPS, v2.4;][]{Conroy2009, Conroy2010} and a \citet{Chabrier2003} IMF over the mass range 0.1--100~$\msun$ are adopted. The Bayesian priors used to fit the photometry are similar to those adopted in \cite{Moustakas2013}, but ``delayed'' star formation histories of the form ${\rm SFR}(t) \propto t\exp(-t/\tau)$ with a uniform prior on $\tau$ in the range $0.1-5$~Gyr are assumed instead of simple exponentially declining star formation histories. We also include nebular emission lines whose strength is tied self-consistently to the number of hydrogen-ionizing photons in the SED of the composite stellar population. In this paper, $M$ refers to the stellar mass resulting from SED-fitting.

\begin{figure}
\centering
\includegraphics[width=0.9\columnwidth]{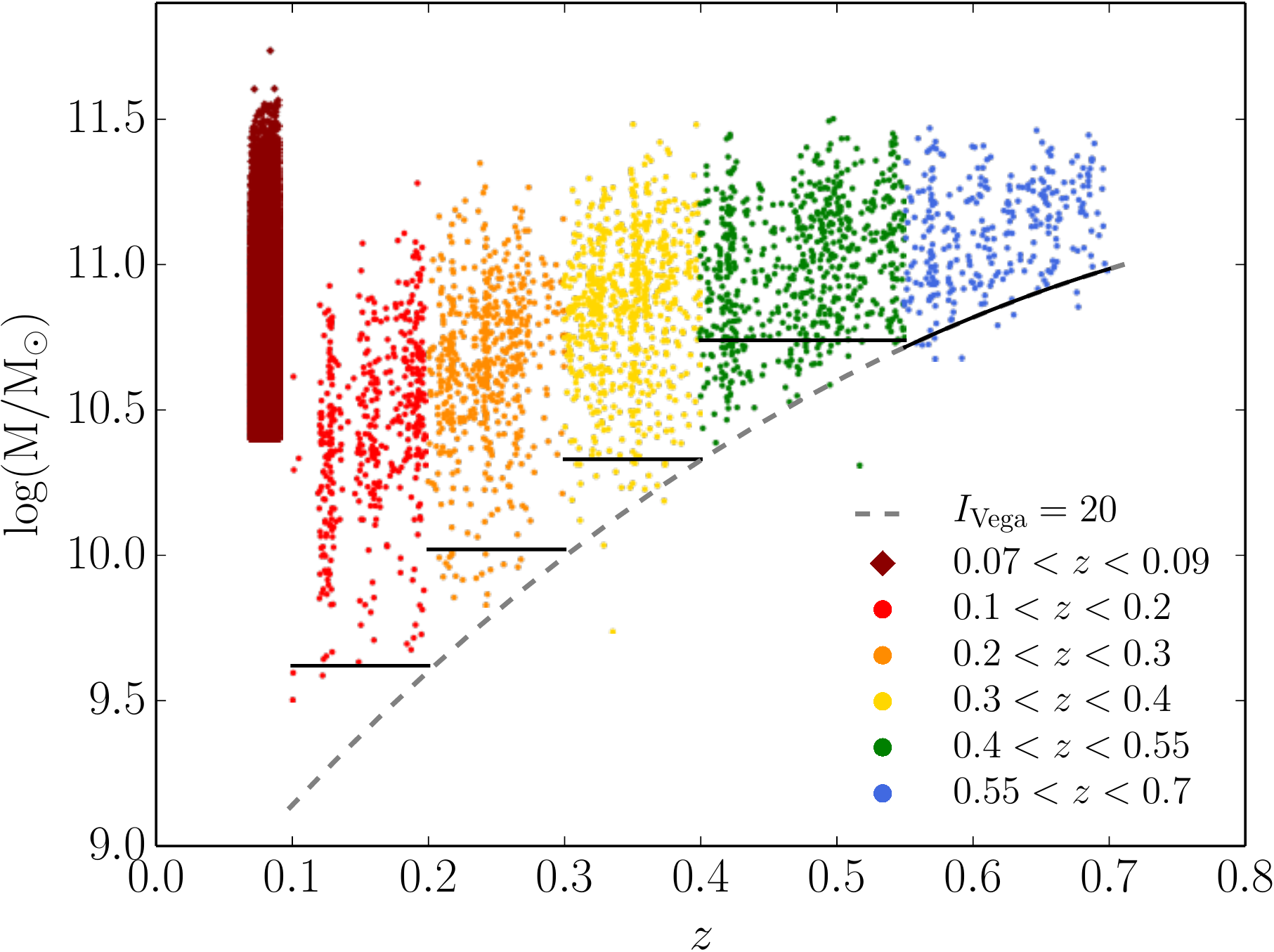}
\caption{Stellar mass as a function of redshift for the quiescent galaxy sample. A subset of the $0.07<z<0.09$ SDSS sample is shown in maroon diamonds and the entire AGES sample is shown in colored circles. The black lines show the mass cuts applied to each color-coded redshift bin and the dashed gray line corresponds to $I_{\rm Vega}$=20 estimated from a photometric redshift sample \citep{Brodwin2006, Brodwin2013}. Our sample is complete in mass in each redshift bin. For $0.55<z<0.7$, all objects above the dashed line are included in order to sample a greater range in stellar mass. Of the resulting three mass bins in this redshift interval, the lowest mass bin is thus technically incomplete. The vertical streaks in the AGES data arise due to the imprints of large scale structure on the relatively small volume covered by the survey.}
\label{fig:AGES_SDSS_mass_redshift}
\end{figure}

From the parent sample, we make the following initial cuts: $100<\sigma\;$(\kms) $<350$, $0.07<z<0.09$, and \mbox{$10.4 < \log(M/\msun) < 11.8$}. The velocity dispersion cut is used to remove objects that are unlikely to be bona fide quiescent galaxies, whereas the redshift selection is a trade-off between overlapping with AGES redshift and reaching a sufficiently low stellar mass limit. The lower mass limit is chosen such that the sample is complete down to that limiting mass. Next, we use specific star formation rate (sSFR) to define the quiescent sample. When viewed in the sSFR versus stellar mass plane, the remaining sample separates into two groups, one that makes up the star-forming main sequence and another that lies below consisting of a quiescent subpopulation. The left panel in Figure~\ref{fig:sSFR_vs_mass} shows a subset of the data, where the two distinct clouds can be clearly seen. To obtain the quiescence cut, the sample is divided into three mass bins---$10.4 < \log(M/\msun) < 11.45$ in bin widths of 0.35\footnote{The highest mass bin is excluded due to its insufficient number of galaxies for the construction of a well-sampled histogram, which is necessary to cleanly locate the bimodality minimum.}---and histograms of sSFR values are constructed. The resulting distributions are clearly bimodal, and we locate the valley and the sSFR value at which it occurs. Next, we perform a linear least-squares fit to these three sSFR values. This line is used to separate the quiescent population from the star-forming main sequence. The resulting quiescence cut is shown as a dashed line in Figure~\ref{fig:sSFR_vs_mass}. The redshift interval for this particular sample is narrow enough that we do not need to consider a cut that evolves with redshift. The final quiescent SDSS sample is shown in Figure~\ref{fig:AGES_SDSS_mass_redshift}.

Finally, we divide the remaining $\sim$37,000 objects into four bins in mass with log-uniform spacing, ranging from $10^{10.4}~\msun$ to $10^{11.8}~\msun$. In order to stack spectra with different continuum shapes and normalizations as well as varying degrees of smoothness due to velocity broadening, we divide each spectrum through by an $n=6$ polynomial to continuum-normalize, then convolve with a Gaussian kernel to achieve an effective dispersion of 350 \kms, the highest $\sigma$ in our entire sample. The goal of the polynomial division is to remove the broad continuum shape while preserving the integrity of the individual absorption features. We have tested the effects of varying the polynomial order and have found that $n$ between 4 and 7 satisfactorily eliminates the broadband shape without significantly affecting the absorption features. The systematic effects of smoothing with a Gaussian kernel are explored in Appendix~\ref{section:smoothing_test}.

Continuum-normalized and smoothed spectra are then coadded with weights provided by the flux uncertainty at each pixel ($w_{\lambda} = 1/{\sigma_{\lambda}}^2$). Each galaxy contributes almost equally to the stack, but a luminous galaxy has smaller Poisson error at every pixel and thus more weight in the stack. As discussed in Section~\ref{section:discussion}, this weighted stacking scheme does not introduce a significant bias toward a small subset of the sample. In other words, the resulting stacks are not dominated by a few bright, young galaxies. The effects of stacking are investigated in more detail in Appendix~\ref{section:stacking_test}, but in summary, the best-fit parameter measured from the stacked spectrum and the weighted average of best-fit parameters from fitting individual spectra agree to within 0.05~dex. Additionally, this weighting scheme serves to mask out bad regions in the spectrum by ensuring zero or very weak contribution from spurious counts with large measurement uncertainties. We fit the stacked spectrum in the rest-frame wavelength range \mbox{4000~--~5500~\aa{}} only in order to ensure identical wavelength coverage across all redshifts for both this low-redshift SDSS sample and the higher redshift AGES sample (see Section~\ref{section:ages_sample}).

\subsection{AGES}
\label{section:ages_sample}
AGES obtained optical spectroscopy for approximately $18,000$ galaxies over 7.7~deg$^2$ of the NOAO Deep Wide-Field Survey (NDWFS) Bo\"{o}tes field \citep{Jannuzi1999, Kochanek2012}.  The survey used the Hectospec instrument \citep{Fabricant2005} on the MMT to obtain 3700~--~9200~\aa{} spectroscopy at a spectral resolution of 6~\aa{} (\mbox{$R\approx1000$}), and achieved a spectroscopic completeness of approximately $90\%$ \citep{Kochanek2012, Cool2012}.  The median redshift of the galaxies in the survey is \mbox{$\langle z\rangle\approx0.3$}, spanning the range $0\lesssim z \lesssim0.8$.  From the full spectroscopic dataset, we select a statistically complete sample of $10,839$ galaxies with well-calibrated spectrophotometry, \mbox{$15<I_{\rm Vega}<20$}, and $0.05<z<0.75$.  

Following \citet{Moustakas2011}, 12 bands of photometry are assembled for this sample, including near- and far-UV photometry from {\it GALEX}/GR6, $u$-band photometry from the LBT/LBC \citep{Bian2013}, {\it $B_{\rm W}RIz$} optical photometry from the NDWFS third data release and zBo\"{o}tes survey \citep{Jannuzi1999, Cool2007}, near-infrared {\it JHK} photometry from a NOAO Extremely Wide-Field Infrared Imager \citep[NEWFIRM;][]{Autry2003} survey\footnote{{\url http://archive.noao.edu/nsa/NEWFIRM\_NDWFS.html}} of the Bo\"{o}tes field, and [3.6] and [4.5] photometry from the \emph{Spitzer} Deep Wide-Field Survey \citep[SDWFS;][]{Ashby2009}. Accurate galaxy colors are measured within a $4\arcsec$ diameter aperture from PSF-matched imaging using a custom code, and tied to the $I$-band {\sc MAG\_AUTO} magnitude measured using {\tt SExtractor} \citep{Bertin1996}. Additional details regarding the photometry can be found in \citet{Brown2007} and \citet{Moustakas2011}. In particular, using simulated galaxies inserted into the NDWFS $I$-band imaging, \citet{Brown2007} showed that {\sc MAG\_AUTO} yields the total galaxy flux within $\approx5\%$ for galaxies brighter than $I_{\rm Vega}=20$ over a wide range of apparent size and galaxy type (i.e., surface brightness profile). Finally, the velocity dispersions for this sample are measured as described in \citet{Moustakas2010, Moustakas2011} using a modified version of the {\textsc pPXF} continuum-fitting code \citep{Cappellari2004}.

Stellar masses and SFRs for AGES are determined by fitting the photometry using {\tt iSEDfit} (see Section~\ref{section:sdss_sample}). After removing galaxies with poorly constrained velocity dispersions ($\lesssim6\%$ of the quiescent sample\footnote{These objects preferentially have low S/N but are uniformly distributed in $U-V$ color within each mass--redshift bin. The omission of these objects from the stacks is thus unlikely to significantly bias the results.}), the following cuts are applied---$100<\sigma\;$(\kms) $<350$ and \mbox{$9.5 < \log (M/\msun) < 11.5$}---leaving approximately $7000$ galaxies. Finally, we select a sample of quiescent galaxies adopting the same sSFR($\log M$) cut derived from the SDSS sample (see the right panel in Figure~\ref{fig:sSFR_vs_mass}). The quiescence cut used here is constant with redshift because there is no evidence for evolution in location of the green valley, i.e., the minimum in the sSFR bimodality.

\begin{figure}[!t]
\centering
	\subfigure{
	\includegraphics[width=0.95\columnwidth]{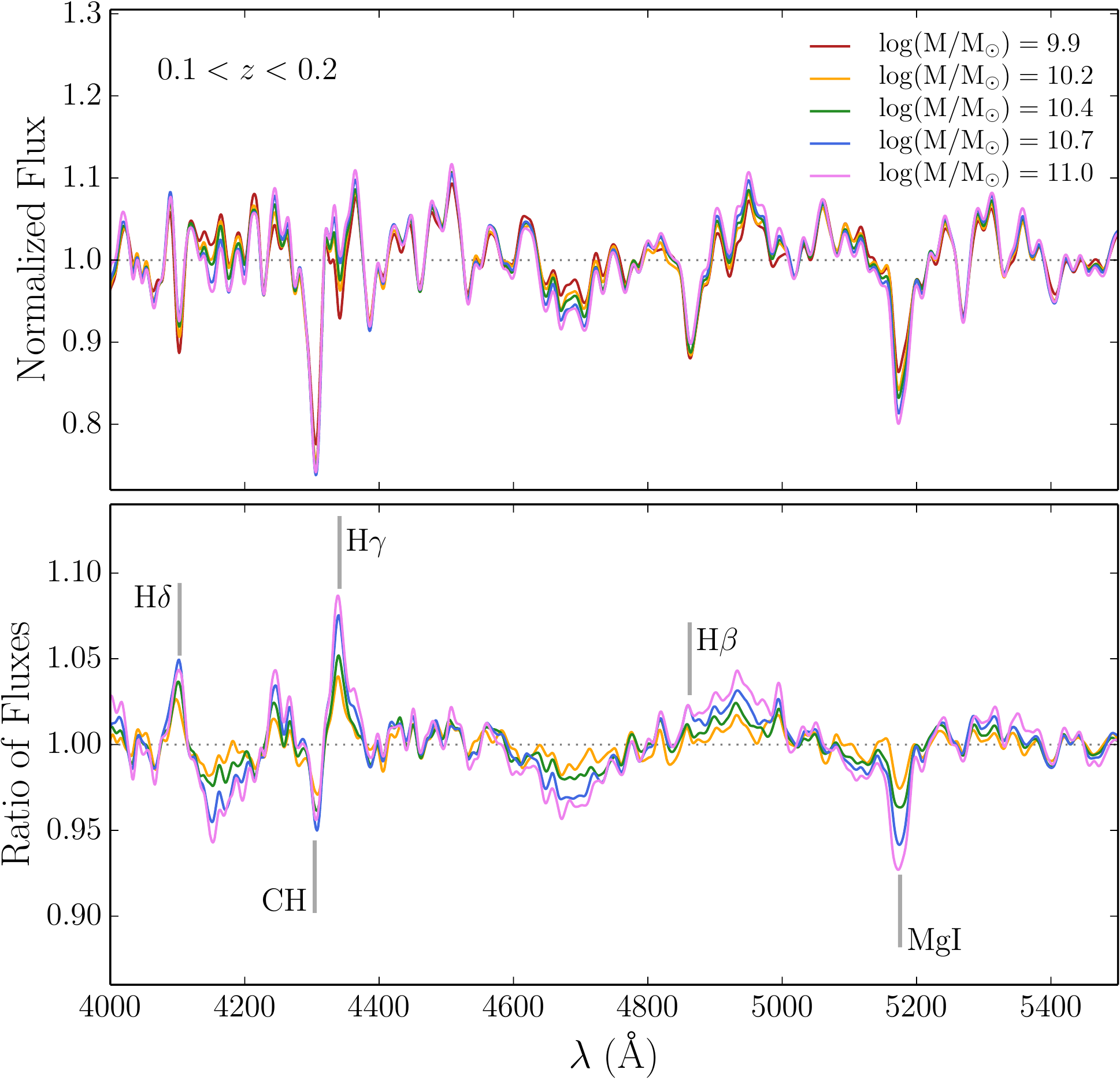}
	}
	\subfigure{
	\includegraphics[width=0.95\columnwidth]{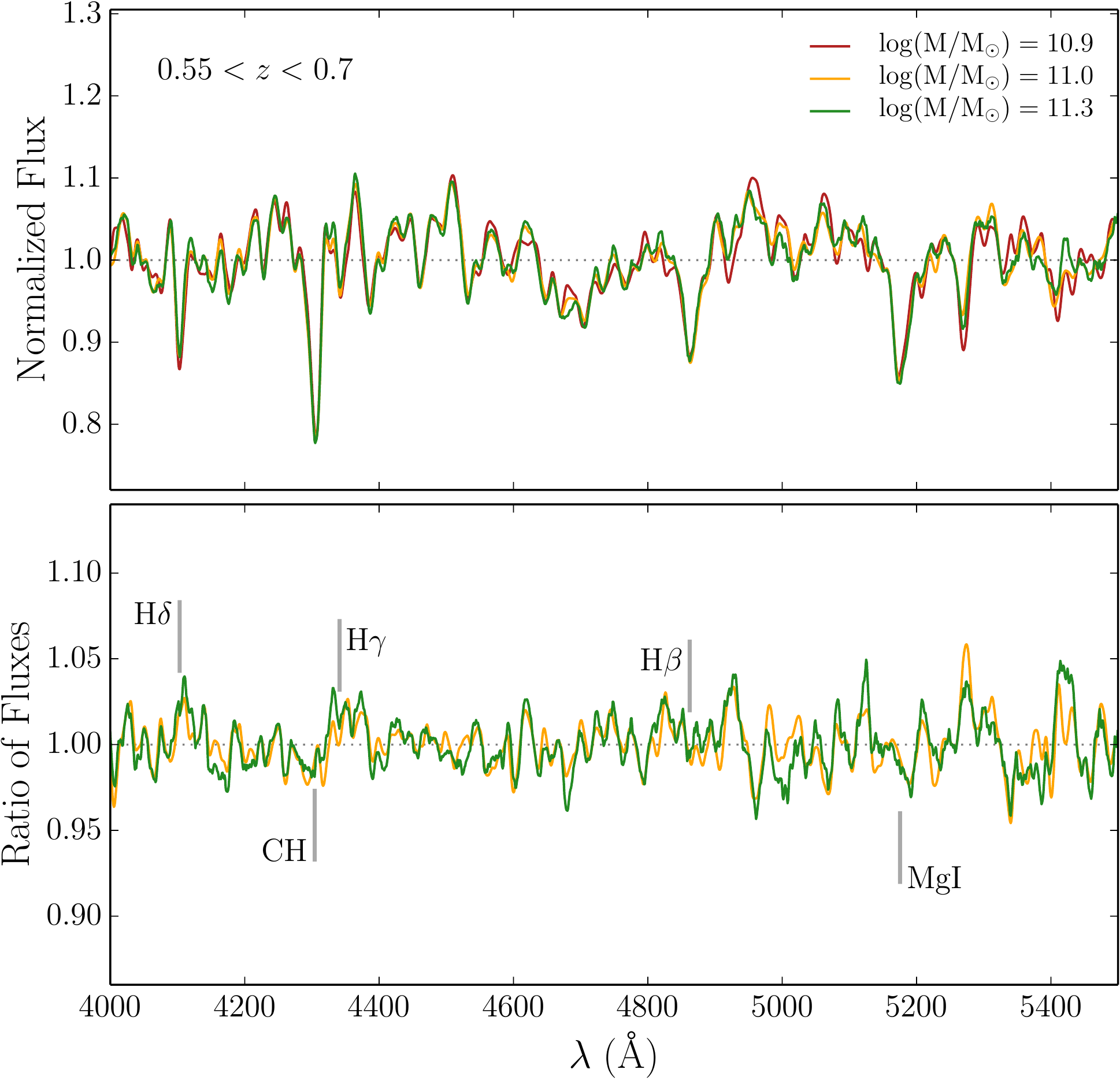}
	}
\caption{Continuum-normalized stacked spectra of AGES quiescent galaxies. Stacked spectra for the lowest and highest redshift intervals are shown in the top panels. The lower panels show the ratio of the fluxes to the flux in the lowest mass bin in each redshift interval using the same color scheme as in their respective top panels. The displayed stellar masses are median values of individual galaxies in each respective bin.}
\label{fig:ages_stacks}
\end{figure}

Since the AGES sample covers a large redshift range, we further partition the data in redshift with the bin divisions at 0.1, 0.2, 0.3, 0.4, 0.55, and 0.7. These uneven redshift slices are meant to reflect roughly constant intervals in time of \mbox{1 Gyr}.

To ensure that the redshift bins are complete in mass, we compute the limiting stellar mass corresponding to $I_{\rm Vega}=20$ using a different, larger photometric redshift sample to compare against our spectroscopic dataset. We utilize the sample from \cite{Brodwin2013}, which updated the photometric redshifts from \cite{Brodwin2006} with deeper \emph{Spitzer} photometry. The photometric redshifts for the 4.5~$\mu$m selected galaxy sample in the $\sim10$~deg$^2$ SDWFS \citep{Ashby2009} are measured from multi-wavelength photometry combining the deep \emph{Spitzer} mid-IR photometry with optical photometry from the NDWFS \citep{Jannuzi1999} and NIR photometry from the FLAMINGOS Extragalactic Survey \citep[FLAMEX;][]{Elston2006}. The stellar masses and SFR are estimated from {\tt iSEDfit}, and quiescent galaxies are selected using a bimodality cut similar to the one applied to the fiducial SDSS and AGES sample. Next, the typical stellar mass corresponding to $I_{\rm Vega}=20$ is calculated via linear least-squares fitting of stellar mass versus $I_{\rm Vega}$ in narrow redshift intervals. The resulting mass limit is used as a guideline to apply completeness cuts to the AGES quiescent galaxy sample, leaving approximately 2400 objects in the final sample. However, for $0.55<z<0.7$, all objects above the $I_{\rm Vega}=20$ limit are included in order to sample a greater range in stellar mass. Of the resulting three mass bins in this redshift interval, the lowest mass bin is thus incomplete and biased toward the lower limit of the redshift bin. 

Figure~\ref{fig:AGES_SDSS_mass_redshift} shows stellar mass as a function of redshift after the quiescence cut has been applied, with the AGES data points shown in colored circles and the black lines demarcating the mass cuts. The dashed gray line shows the $I_{\rm Vega}=20$ limit from the photometric redshift sample. The vertical streaks, especially prominent in the $0.1<z<0.2$ slice, are a consequence of sampling a relatively small volume of the large scale structure: the survey cone happened to intersect three overdense regions within which the galaxies in our sample are clustered.

We divide each redshift interval into six mass bins, continuum-divide, smooth the spectra to an effective velocity dispersion of 350 \kms, and stack the spectra. See Section \ref{section:sdss_sample} for more detail on this stacking process. We do not bin our spectra in velocity dispersion because a given velocity dispersion bin is incomplete in mass at the highest redshifts. There are tens to hundreds of objects in each mass--redshift bin, but one of them---corresponding to $0.2<z<0.3$ and $11.2< \log (M/\msun) < 11.5$---contains only five objects. More details regarding the stacked spectra can be found in Table~\ref{table:results}.

In Figure~\ref{fig:ages_stacks} we show the stacked AGES spectra in several mass bins for both the lowest and the highest redshift intervals to highlight the differences in their spectral features. The bottom panels show the ratio of the fluxes to the flux in the lowest mass bin in each redshift interval. The typical S/N for the low- and high-redshift stacked spectra are $\sim100\text{--}200$~\AA{}$^{-1}$ and $\sim50$~\AA{}$^{-1}$, respectively. While differences are fairly modest overall, variations in the strengths of absorption features such as H$\delta$ at 4103~\aa{}, CH at 4300~\aa{}, H$\gamma$ at 4341~\aa{}, H$\beta$ at 4862~\aa{}, and MgI at 5175~\aa{} are clearly noticeable.

\subsection{Keck DEIMOS}
In addition to the two survey data sets, we also analyze DEIMOS \citep{Faber2003} spectra of the two brightest quiescent galaxies in the magnitude-limited survey of cluster MS 1054-03 at $z=0.83$ \citep{Holden2010}. They were observed in 2 hr exposure intervals using the 600 line mm$^{-1}$ grating for a total integration time of 20 hr to achieve a \mbox{S/N $\approx$ 30 \aa$^{-1}$} at 5000~\AA{}. Additional details about the two objects, 7415 and w5756, can be found in Table A1 in \citet{Holden2010}. These spectra are fit individually from 4000~\aa{} to 5440~\aa{} after continuum-normalization.

\section{MODEL FITTING}
\label{section:model}
The main analysis tool for our work was developed by \cite{Conroy2012} and subsequently updated in \cite{Conroy2014}. The technique involves full spectrum fitting of optical-NIR spectra (3500~\aa~--~2.4~$\mu$m with $R \sim 2000$) of quiescent galaxies. It is an alternative method to the traditional technique of modeling the Lick/IDS indices \citep{Burstein1984, Worthey1994, Trager1998, Thomas2005, Schiavon2007}. A defining feature of this code is that it fits the entire continuum-normalized spectrum. The main advantage of this technique is that since all of the absorption features are fit simultaneously, there is more information available to better constrain the fit, resulting in a reliable measurement of the stellar population parameters even at low S/N (see Appendix~\ref{section:snr_test}). This implies that for a fixed S/N per \aa{}, the age-metallicity degeneracy \citep{Worthey1994} is less pronounced when fitting the full spectrum (see also \citealt{Sanchez2011}).

\begin{figure}[!t]
\centering
\includegraphics[width=0.9\columnwidth]{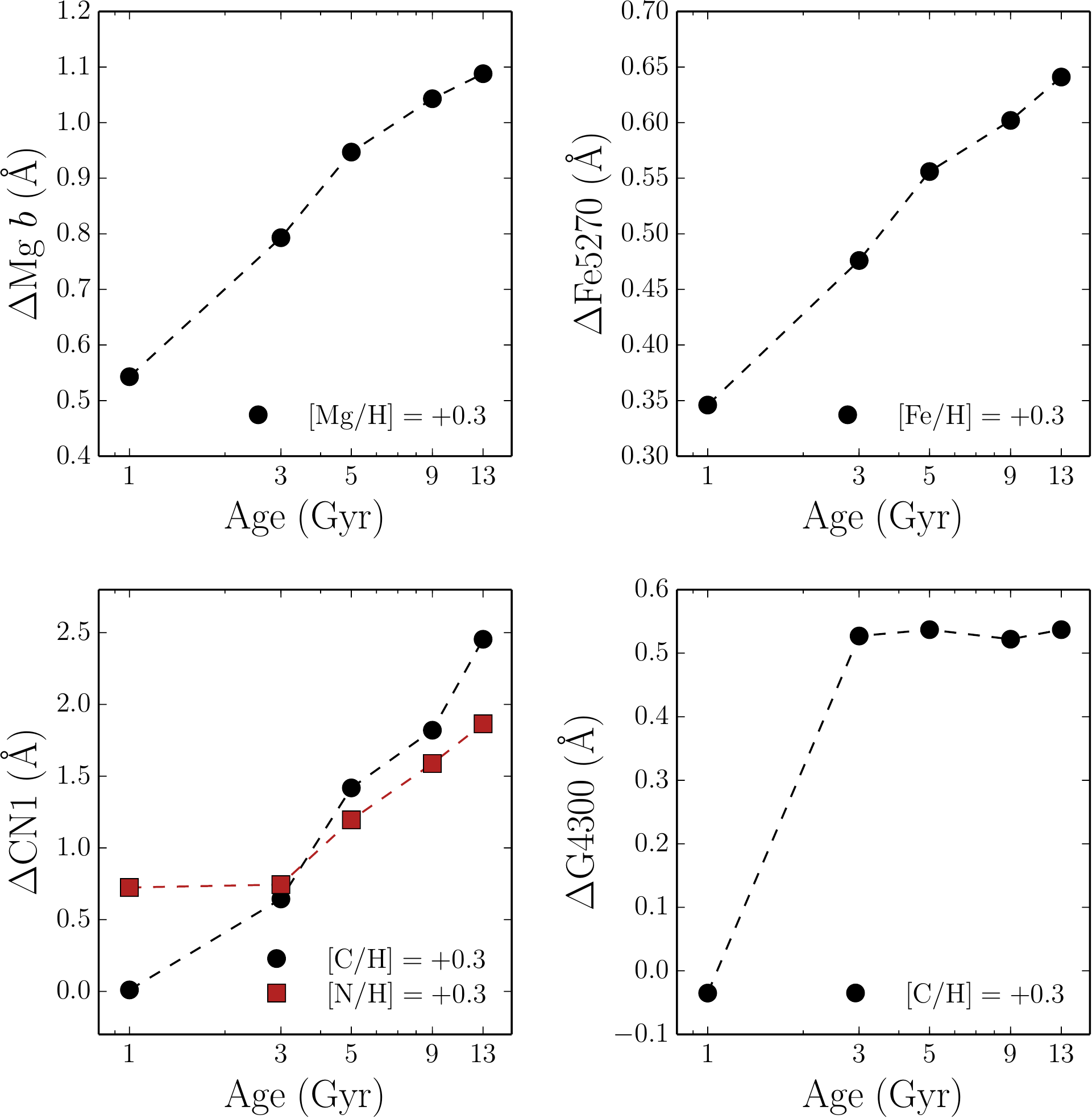}
\caption{Change in Lick indices as a function of age due to a 0.3~dex increase in various elements. These panels highlight the importance of using age-dependent response functions when modeling the spectra of quiescent galaxies. The G4300 index measures the strength of the CH band at 4300~\aa{}. Note that the indices shown here are for illustrative purposes only. In our main analysis we fit models to the full optical spectra.}
\label{fig:responsefunc_index}
\end{figure}

As described in \cite{Conroy2012} and \cite{Conroy2014}, the model follows standard stellar population synthesis techniques by making use of stellar isochrones and empirical stellar spectral libraries to fit the wavelength range 0.35--2.4~$\mu$m. Since the stars used to construct the spectral libraries have solar abundance patterns and are roughly $Z \sim Z_{\odot}$, necessary adjustments must be made to account for $\alpha$-enhancements and super-solar metallicities observed in quiescent galaxies. To achieve this task, the code makes use of response functions constructed using a large grid of model stellar atmospheres and spectra computed with the ATLAS12 package \citep{Kurucz1970, Kurucz1993}, ported to Linux by \cite{Sbordone2004}. These response functions \citep[see, e.g., Figure 2, 3, 4 in][]{Conroy2014} are derived for each element as a function of wavelength, which are then used to modify the template model galaxy spectrum. The model first presented in \cite{Conroy2012} and subsequently updated in \cite{Conroy2014} has since been updated to include younger populations. Whereas the first generation model ranged from 3--13.5~Gyr in age, the model is now capable of fitting populations as young as 1~Gyr. We note that the stellar isochrones are not recomputed for different elemental mixtures in the model. To account for a shift in effective temperature of the isochrone due to changes in the abundance patterns, \cite{Conroy2012} included $\Delta T_{\rm eff}$ as an additional parameter in the fit. In the `simple' version of the code used in the present work, we do not include $\Delta T_{\rm eff}$. This is not expected to cause an appreciable systematic effect \citep{Conroy2012}, though this will be tested in future work. We also note that this assumption is standard in the modeling of old stellar populations \citep{Thomas2005, Schiavon2007}.

\begin{figure}[t!]
\centering
	\subfigure{
	\includegraphics[width=0.9\columnwidth]{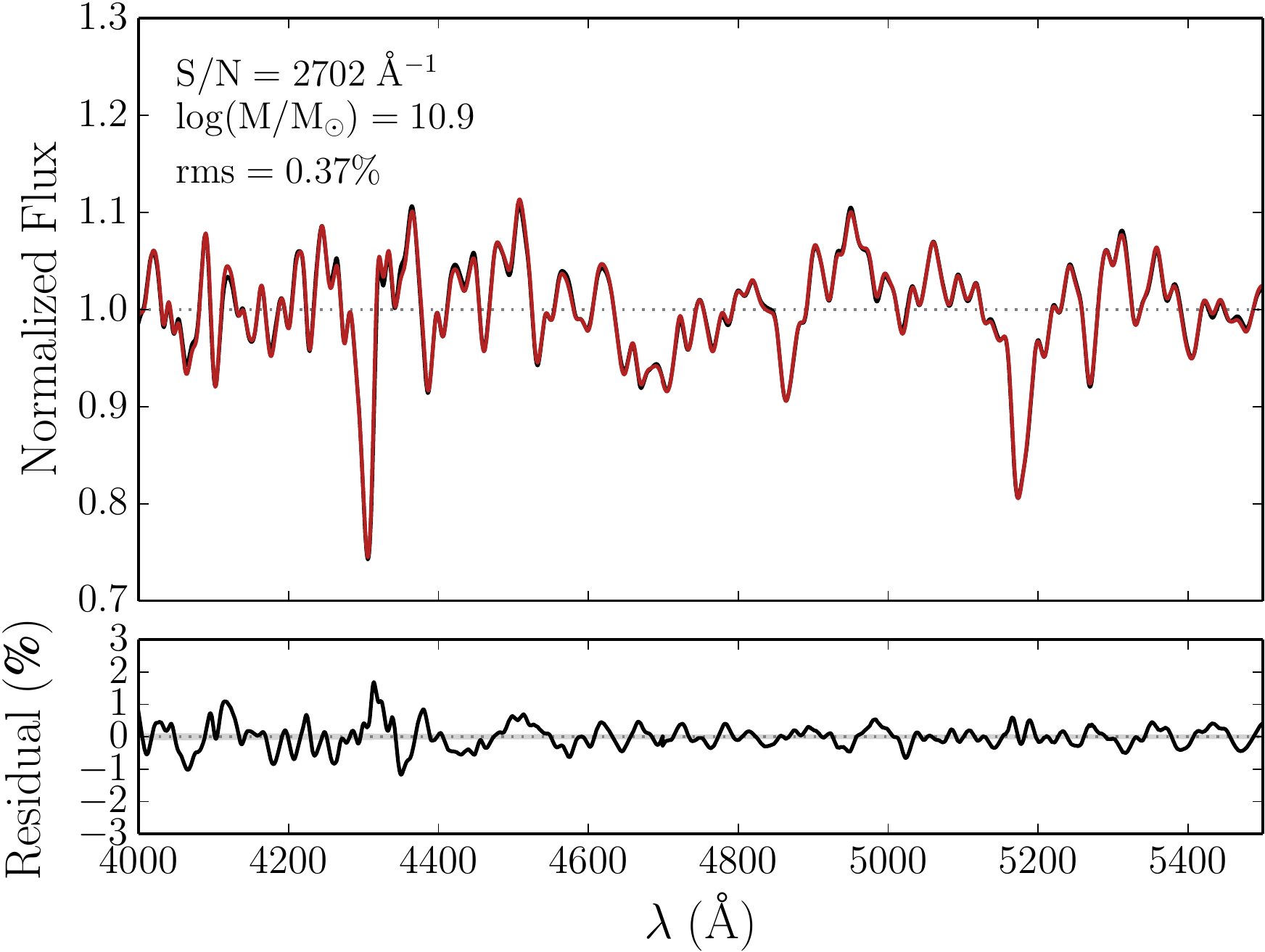}
	}
	\subfigure{
	\includegraphics[width=0.9\columnwidth]{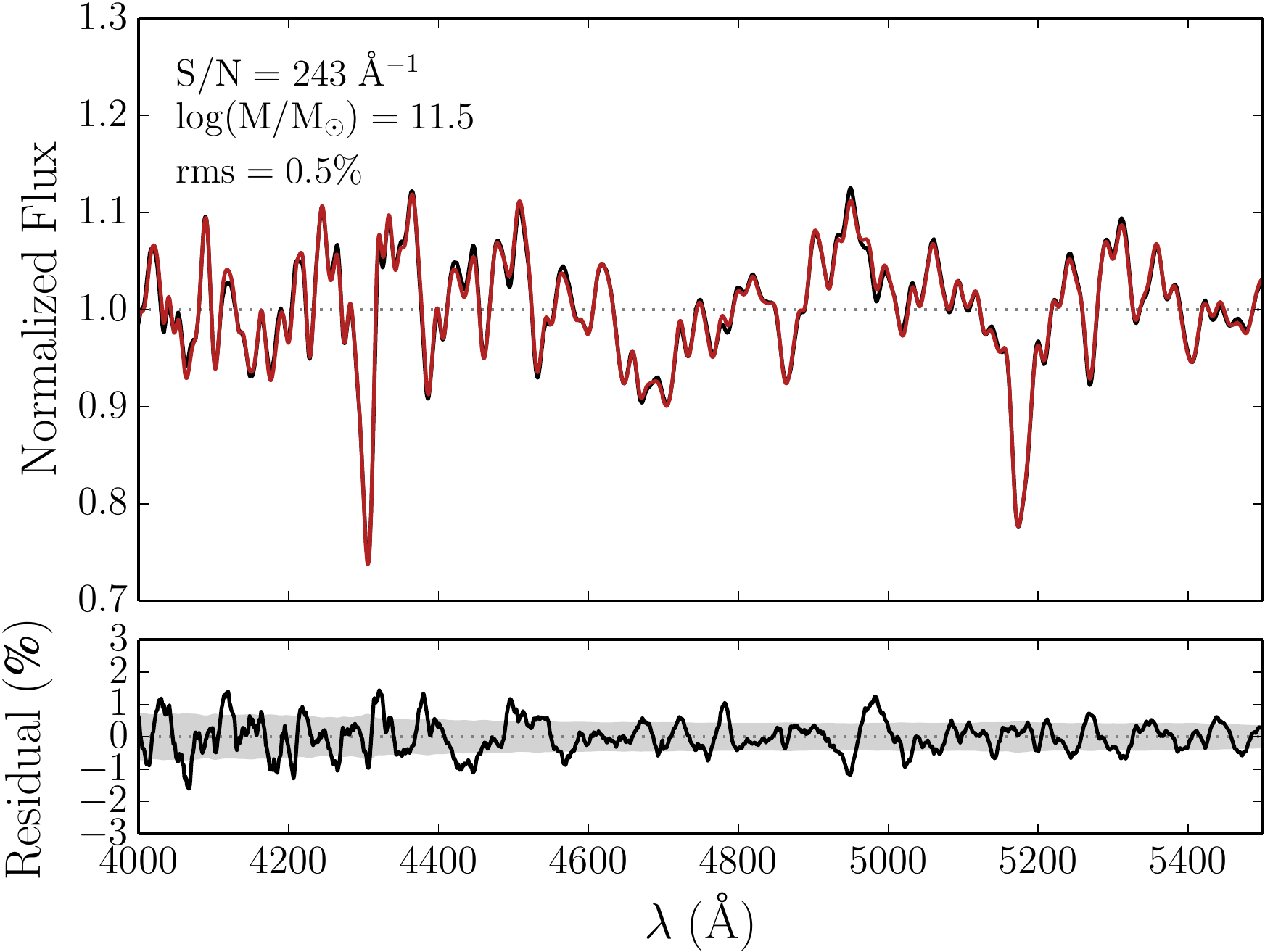}
	}
\caption{Representative comparison between SDSS stacked spectra and the corresponding best-fit models, displayed in black and red, respectively. The flux has been continuum-normalized by a high-order polynomial. The bottom panel shows the fractional residuals in the models in black and the flux uncertainties as the gray shaded region. The residual is formally larger than is allowed by the uncertainties in the data, but the stacked data are of exquisite quality, which imposes very strong demands on the model. The displayed stellar masses are median values of individual galaxies in each respective bin, while the S/N is the median calculated in the wavelength range \mbox{4000~--~5500~\aa.{}}}
\label{fig:sdss_examplefit}
\end{figure}

\begin{figure}[t!]
\centering
	\subfigure{
	\includegraphics[width=0.9\columnwidth]{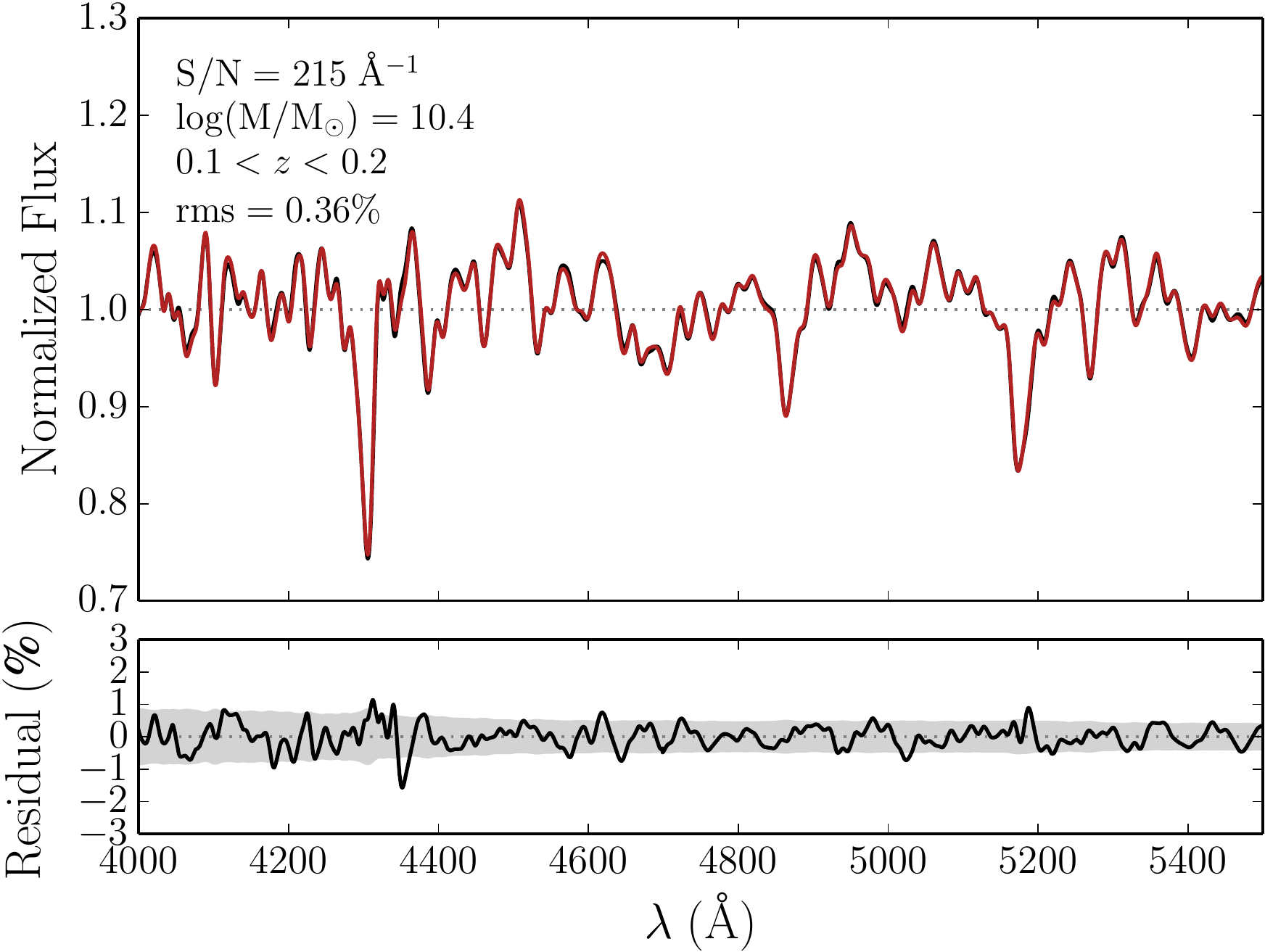}
	}
	\subfigure{
	\includegraphics[width=0.9\columnwidth]{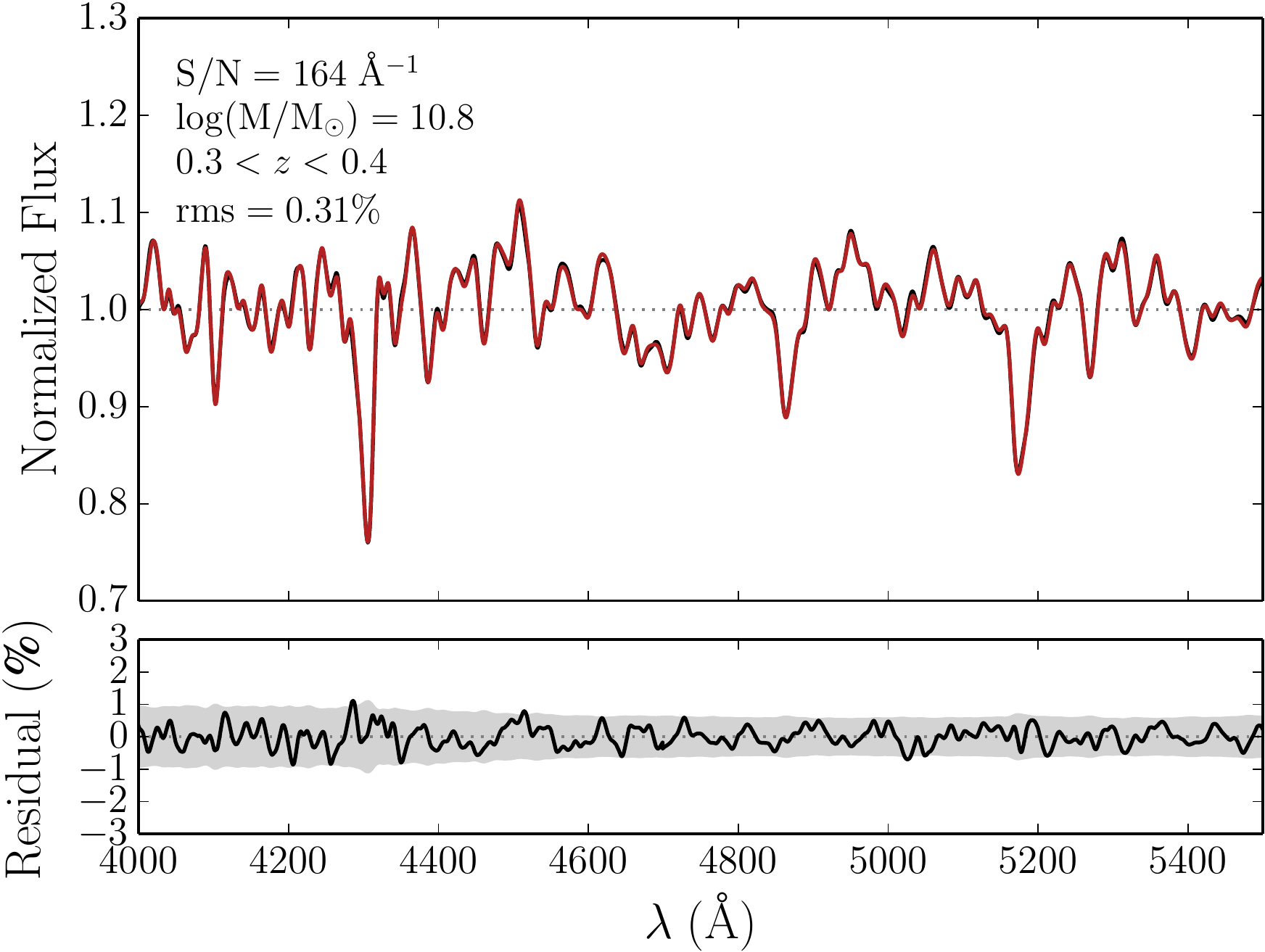}
	}
	\subfigure{
	\includegraphics[width=0.9\columnwidth]{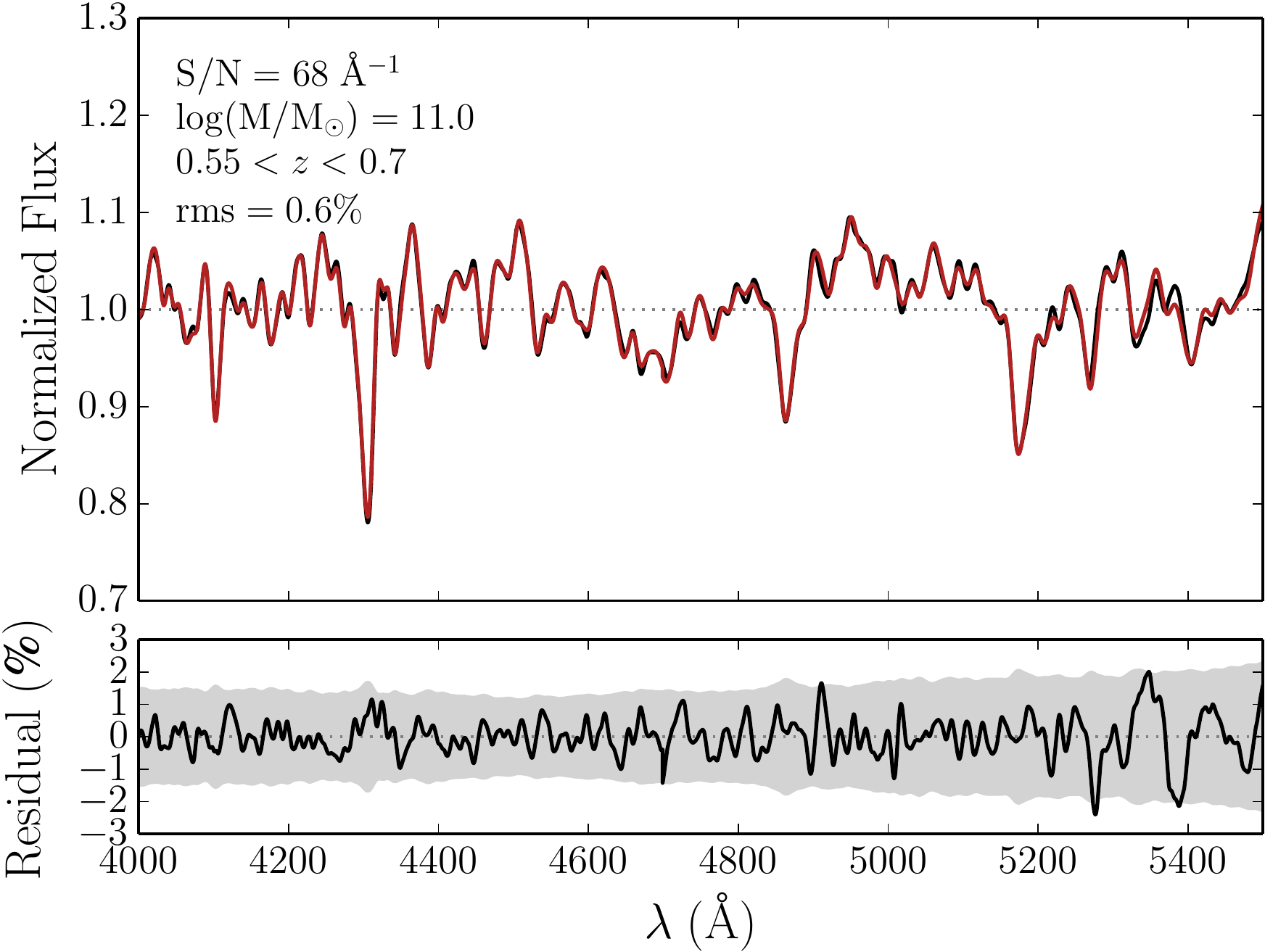}
	}
\caption{The same as Figure~\ref{fig:sdss_examplefit}, now for AGES stacked spectra.}
\label{fig:ages_examplefit}
\end{figure}

\begin{figure}[]
\centering
\includegraphics[width=0.95\columnwidth]{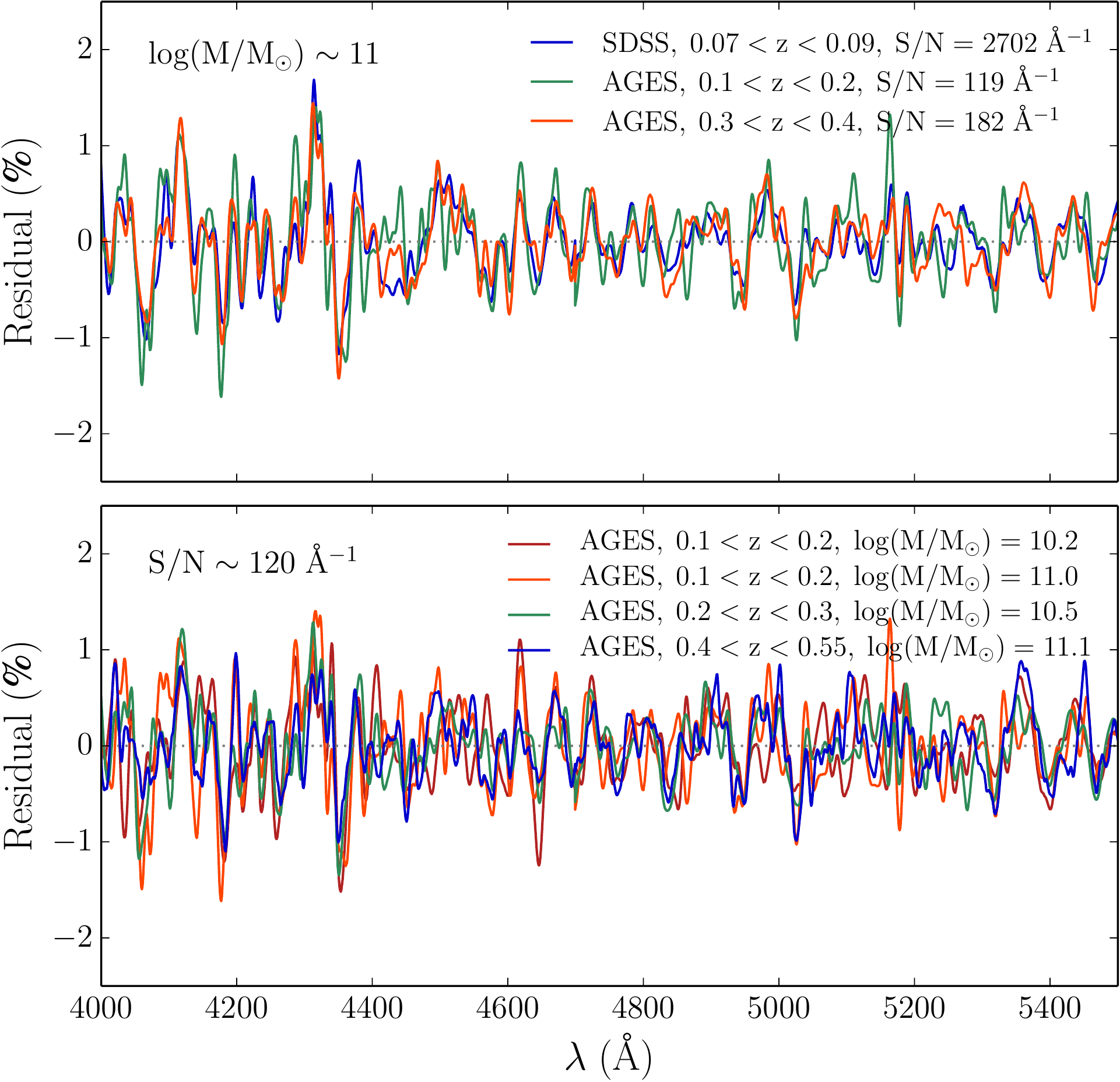}
\caption{Residuals between the data and best-fit models for AGES and SDSS stacked spectra. {\it Top panel: } a comparison between SDSS and AGES $0.1<z<0.2$ and $0.3<z<0.4$ bins, with median masses around $\log (M/\msun) \sim 11$, showing that the residuals are very well-behaved. This demonstrates that any coherent behavior between these residuals is due to the model and not dominated by differences in the data, such as the sampled redshift range, reduction pipeline, or sample selection criteria. {\it Bottom panel: } a comparison between different AGES mass--redshift bins with median S/N around 120~\aa$^{-1}$. The residuals show no systematic trend with mass or redshift, confirming the findings of \cite{Conroy2014} that both high-mass galaxies with $\alpha$-enhanced abundance patterns and low-mass galaxies with roughly solar-abundance patterns are well-described by our model.}
\label{fig:compare_residuals}
\end{figure}

An additional improvement to the model since \cite{Conroy2014} is the inclusion of time-dependent response functions. Originally, the response functions were constructed only for a 13~Gyr isochrone. Given the nature of this current work, however, we have deemed it important to consider the response functions at younger ages (1, 3, 5, and 9~Gyr). For demonstrative purposes, we show the change in Lick indices due to a 0.3~dex increase in various elemental abundances as a function time in Figure~\ref{fig:responsefunc_index}.\footnote{Note that no attempt was made to place these indices on the standard Lick/IDS index zero point scale. The indices are measured on spectra broadened to 100~\kms{}.} All indices are quoted as equivalent widths with units of \AA{} \citep[for detail on index measurements, see][]{Worthey1994}. The indices become less sensitive to changes in abundances at young ages (see also \citealt{Lee2009}). Changes in elemental abundances not only influence the absorption features themselves but the continuum as well. These new response functions will be described in detail in future work.

Exploration of parameter space is achieved using a Markov Chain Monte Carlo (MCMC) algorithm, which yields error estimates on the fit parameters derived from the full posterior distributions, i.e., by marginalizing over the other parameters. These errors are statistical only. A typical systematic uncertainty of $\pm~0.05$~dex should be assumed in addition to the statistical errors \citep[see the Appendix and also ][]{Conroy2014}. The original MCMC algorithm originally used in \cite{Conroy2012} has since been replaced with {\tt emcee}, an implementation of the affine-invariant ensemble sampler for MCMC \citep{ForemanMackey2013}.\footnote{{\url{http://dan.iel.fm/emcee}}} The main advantage of {\tt emcee} is that there are only one to two hand-tuned parameters, a dramatic decrease compared to $\sim N^2$ required for a traditional MCMC algorithm sampling a $N$-dimensional parameter space.

The spectrum fitting code is run in `simple' mode, which measures 10 stellar population parameters---simple stellar population (SSP)-equivalent age,\footnote{We note that light-weighted and SSP-equivalent ages are distinct concepts. Light-weighted age is obtained by integrating over SFR($t$) weighted by luminosity, whereas SSP-equivalent age specifically assumes a single burst. SSP-equivalent ages are generally younger than light-weighted ages, and both are younger than mass-weighted ages. See also \cite{Trager2000b}, \cite{Serra2007}, and \cite{Trager2009}.} [Fe/H], [Mg/Fe], [O/Fe], [C/Fe], [N/Fe], [Na/Fe], [Si/Fe], [Ca/Fe], and [Ti/Fe]---in addition to redshift and velocity dispersion over the wavelength range 4000~--~5500~\aa{}. Since we are not fitting wavelengths redward of 5500~\aa{}, which harbor dwarf-sensitive features, IMF parameters are not explored in the fit and a Kroupa IMF \citep{Kroupa2001} is enforced.\footnote{The difference between stellar masses computed assuming a Kroupa and Chabrier IMF is $<0.05$~dex.} Additionally, [Na/Fe] is set to track [Mg/Fe] because the spectral range we fit does not contain Na-sensitive features, but this assumption has little consequence in our fits. We tested the effects of changing the reddest limit in the wavelength range we fit (varying from 5300~\aa{} to 5800~\aa) and found that the resulting parameters vary comfortably within their 1$\sigma$ errors. We refer the reader to the Appendix for more details regarding additional systematic tests.

\begin{figure*}
\centering
\includegraphics[width=2\columnwidth]{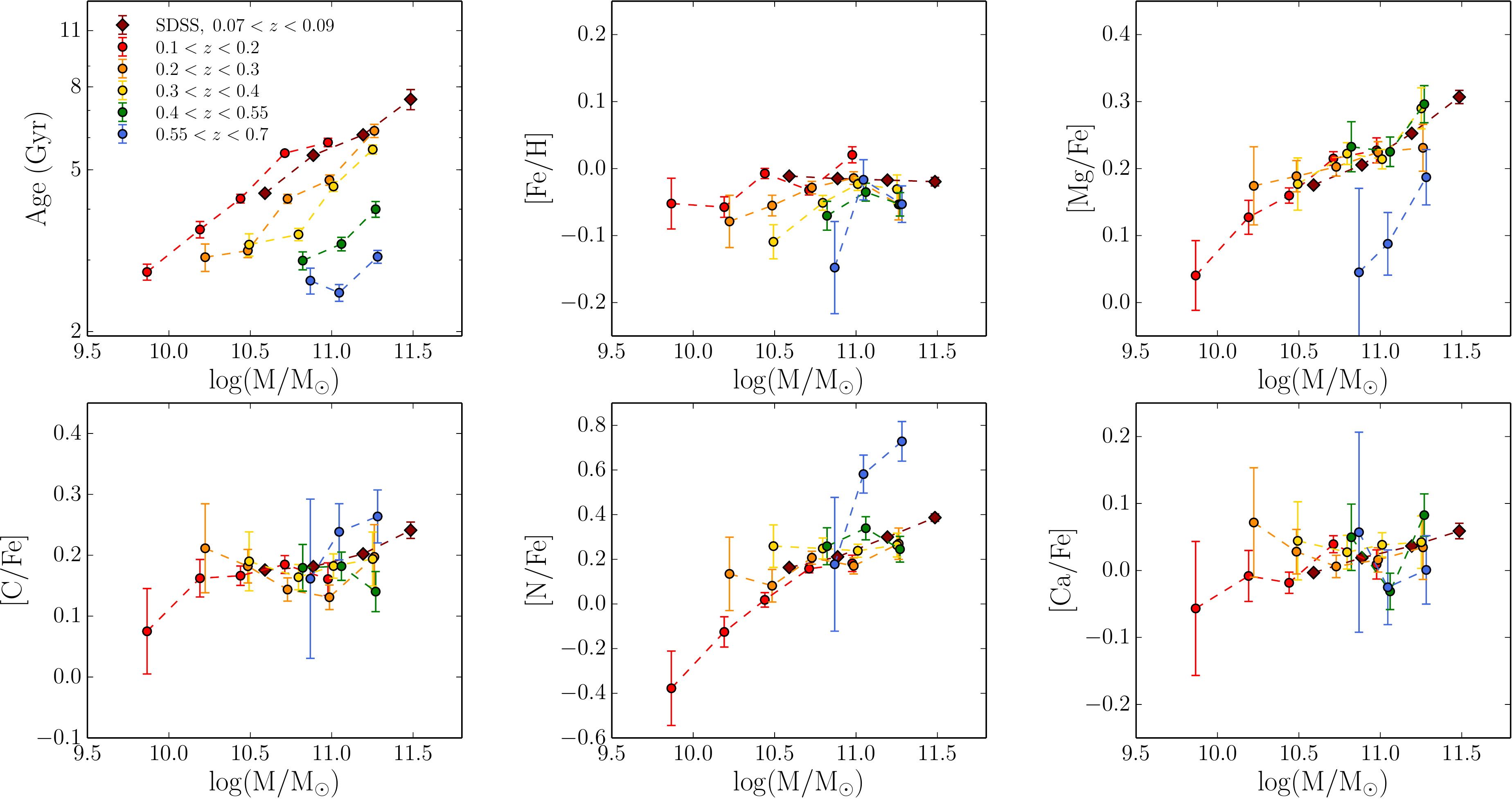}
\caption{Best-fit stellar population parameters shown as a function of stellar mass for SDSS (maroon diamonds) and AGES (red, orange, yellow, green, and blue circles) data. The mass values correspond to the median stellar mass of the samples within each bin. The lowest mass bins are absent at higher redshifts because the parent sample is magnitude-limited and thus does not include distant low-mass galaxies. The error estimates are statistical only and come directly from the Markov Chain Monte Carlo spectrum-fitting algorithm. Note the change in $y$-axis scale between panels. Overall, the abundance patterns show no evidence for evolution with redshift at fixed stellar mass over the last half of cosmic time.}
\label{fig:ages_sdss_mass}
\end{figure*}

\begin{figure}[]
\centering
\includegraphics[width=0.85\columnwidth]{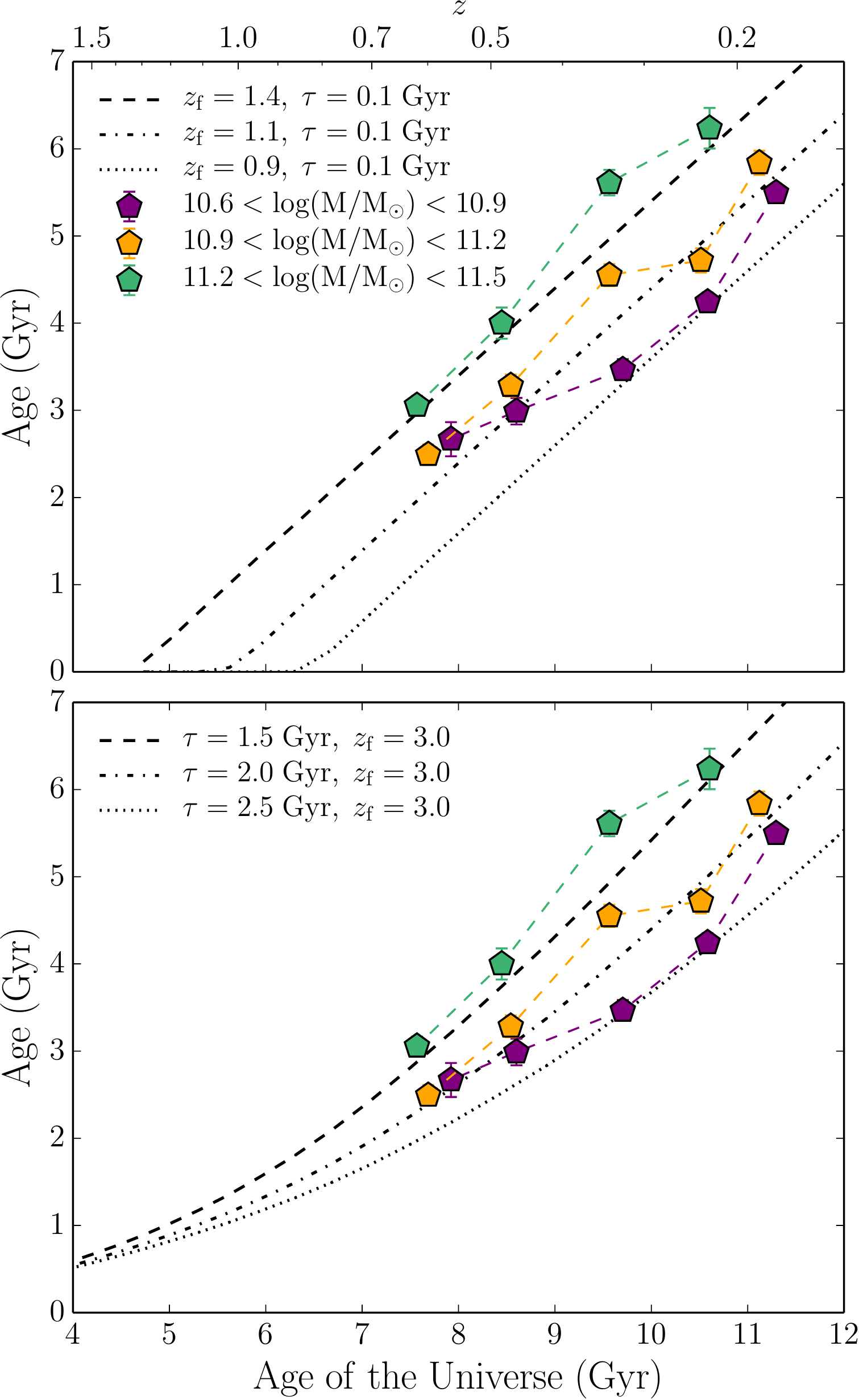}
\caption{{\it Top panel: } age evolution of AGES galaxies for the three highest mass bins. Lower mass bins are not shown because they do not span a wide range in redshift. Each set of colored pentagons corresponds to a single mass bin. The age corresponds to the SSP-equivalent age of the representative galaxy in the mass--redshift bin. The three curves represent the evolution of light-weighted age of models with different star formation histories of the form ${\rm SFR}(t) \propto \exp(-t/\tau)$. The tracks correspond to passive evolution following a burst ($\tau=0.1$~Gyr) at different formation redshifts, $z_{\rm f}$. {\it Bottom panel: } the same as the top panel, but now for the same formation redshift, $z_{\rm f}=3.0$, and varying $\tau$. The age evolution of the most massive galaxies is consistent with passive evolution while there is evidence for a shallower age evolution for the lower mass galaxies shown in this figure. This suggests that the age evolution of the lower mass galaxies is diluted, or slowed down, by the addition of newly-quenched, young galaxies.}
\label{fig:logage_redshift}
\end{figure}

\section{RESULTS}
\label{section:results}
We now present our main science results from modeling the spectra of $z=0.1~\text{--}~0.8$ quiescent galaxies. We first review the $z=0.1~\text{--}~0.7$ AGES and SDSS data together, then introduce the Keck DEIMOS $z=0.83$ cluster data.

\subsection{Stellar Population Parameters of Quiescent Galaxies from AGES and SDSS Spectra}
We begin by showing representative SDSS and AGES spectra and best-fit models in Figures~\ref{fig:sdss_examplefit} and \ref{fig:ages_examplefit}. The continuum-normalized fluxes are shown in black while the models are plotted in red in the top panels.\footnote{We note that the intrinsic variation due to spectral features is $\sim5$\%--10\%, so the residuals with rms of $\sim0.5$\%--1\% correspond to a factor of $\sim10$ improvement in the quality of the fit over a straight line.} By eye, the models are excellent fits to these high S/N stacked spectra. The bottom panels show the percent fractional residuals between the best-fit model and data in black and the flux uncertainties as the gray shaded region. In Figure~\ref{fig:sdss_examplefit}, the fit is worse than is formally allowed by the uncertainties in the SDSS data, but the S/N is extremely high, which imposes strong demands on the model. In Figure~\ref{fig:ages_examplefit}, the residuals are comfortably within the data uncertainties. The displayed $\log M$ values are the median values of individual galaxies in each respective bin, while the S/N is the median calculated in the wavelength range 4000~--~5500~\aa.

In Figure~\ref{fig:compare_residuals} we compare the residuals for AGES and SDSS stacked spectra. The top panel shows a comparison between SDSS and AGES $0.1<z<0.2$ and $0.3<z<0.4$ bins, with median masses around $\log (M/\msun) \sim 11$. The residuals are very well-behaved, demonstrating that any coherent behavior between these residuals is due to the model and not dominated by differences in the data, such as the sampled redshift range, reduction pipeline, or sample selection criteria. The size of the fluctuations is comparable despite the wide range in the median S/N values, implying that the residuals eventually hit a floor even when fitting exquisite high S/N spectra. The bottom panel shows a comparison between different AGES mass--redshift bins with median S/N around 120~\aa$^{-1}$. The residuals show no systematic trend with mass or redshift, confirming the findings of \cite{Conroy2014} that both high-mass galaxies with $\alpha$-enhanced abundance patterns and low-mass galaxies with roughly solar-abundance patterns are well-described by our model. In other words, the residuals are present even in the low-mass stacks where the abundance ratios are roughly solar, suggesting that the theoretical templates are not a major sources of the systematic residuals.

Now we present the main science results of this work. In Figure~\ref{fig:ages_sdss_mass} we display the resulting best-fit SSP-equivalent age and abundance measurements from full spectrum fitting of AGES and SDSS stacked spectra as a function of stellar mass. The SDSS and AGES data are shown in maroon diamonds and red, orange, yellow, green, and blue circles, respectively. The stellar masses were estimated from broadband UV, optical, NIR, and mid-IR SEDs with the {\tt iSEDfit} modeling code (see Sections~\ref{section:sdss_sample} and \ref{section:ages_sample}). The displayed mass for each bin corresponds to the median stellar mass of the samples within that bin. Although other elemental abundance ratios---[O/Fe], [Na/Fe], [Si/Fe], and [Ti/Fe]---are included in the fit, the effects of these elements on the spectra are smaller than the other elements, so they are not discussed in this paper. As previously mentioned, Na is set to track Mg, and the response functions for O, Si, and Ti require data at other wavelengths for robust results and/or are generally weaker than the others, requiring higher S/N spectra \citep{Conroy2014}. The results are summarized in Table~\ref{table:results}.

There is good agreement between SDSS and AGES data at $z\sim0.1$. This implies that the spectroscopic pipelines and other aspects of the data reduction for both surveys contribute minimally to the error budget. Additionally, the difference in spectroscopic fiber size (SDSS fiber is twice as large as the Hectospec fiber for AGES) appears to have little effect, which means that there are no discernible radial gradients between 0.3 and 0.7~$R_{\rm e}$. Lastly, as we demonstrate in Appendix~\ref{section:snr_test}, we do not expect the differences in S/N to systematically affect the AGES results relative to the high-S/N SDSS results.

The well-known trend of the most massive galaxies harboring the oldest stellar populations is evident in Figure~\ref{fig:ages_sdss_mass}. As expected, the age increases with decreasing redshift at a fixed mass. There is no strong evidence for the variation of [Fe/H] with either mass or redshift. Overall, [Fe/H] is very slightly sub-solar, and it varies by $\lesssim0.05$~dex over the full mass and redshift range. We find a positive trend between [Mg/Fe] and mass, but no evidence of evolution with redshift.

\begin{figure}[t!]
\centering
	\subfigure{
	\includegraphics[width=0.9\columnwidth]{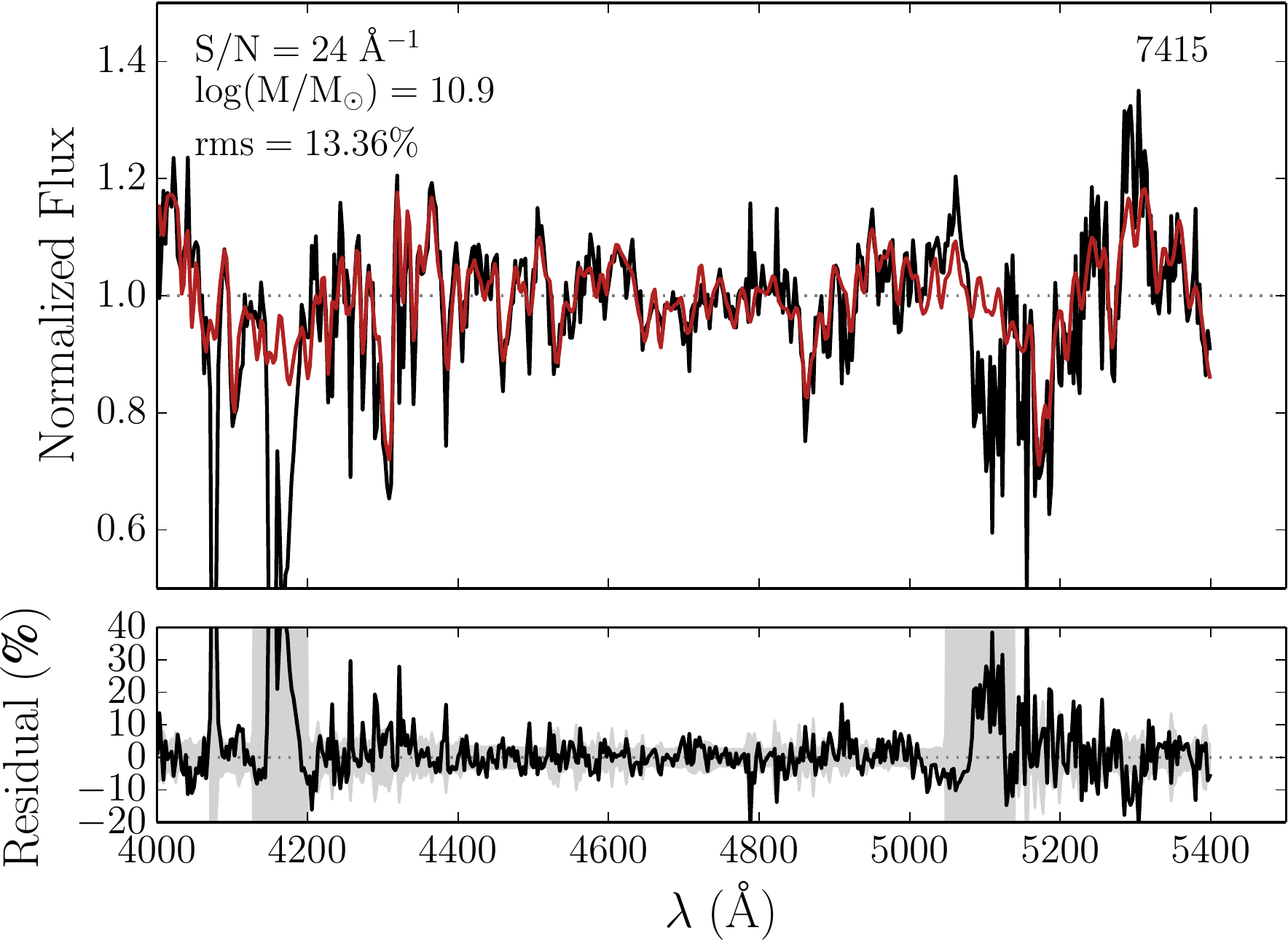}
	}
	\subfigure{
	\includegraphics[width=0.9\columnwidth]{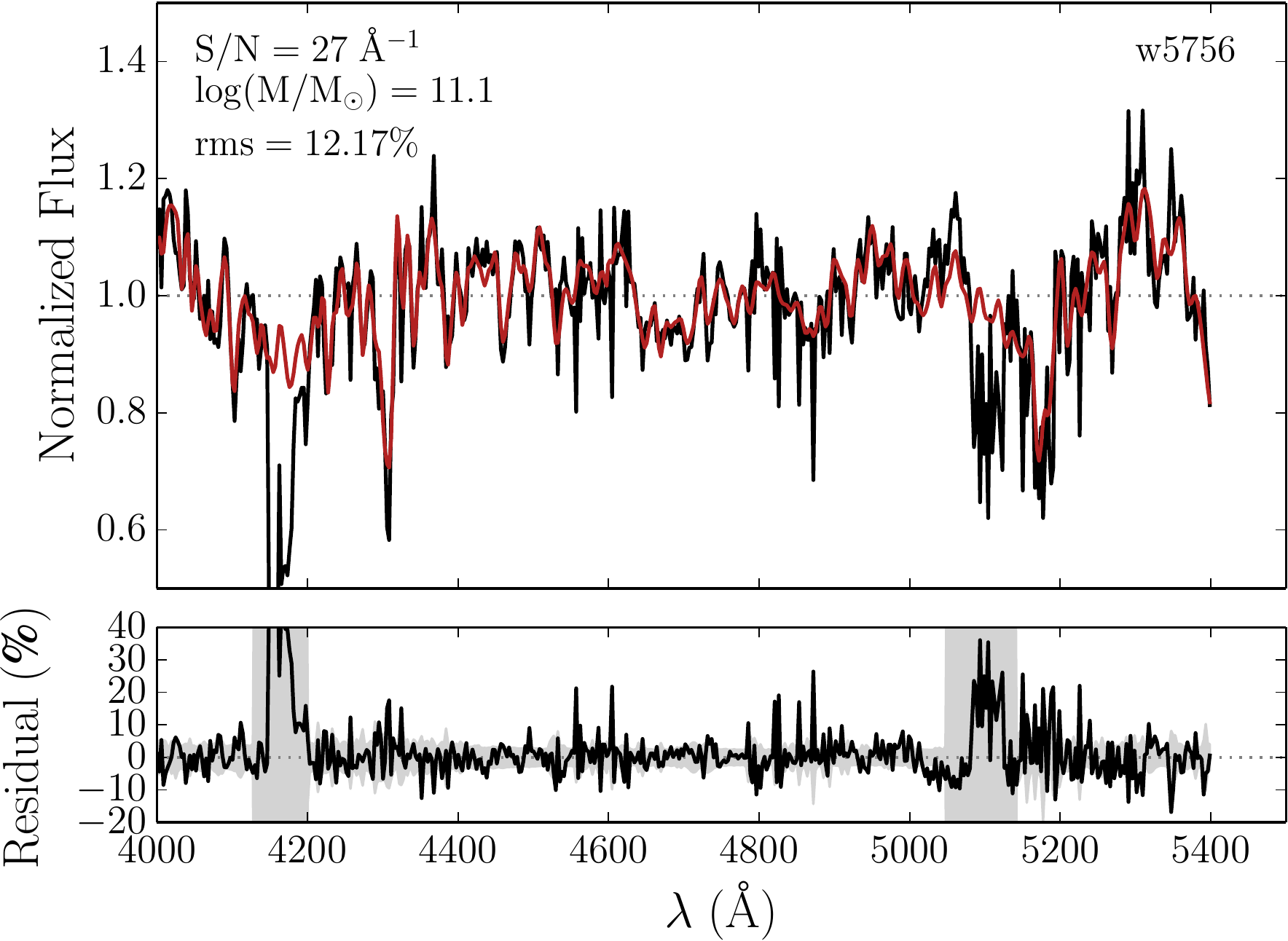}
	}
\caption{Same as Figure~\ref{fig:sdss_examplefit}, now for Keck DEIMOS spectra of the two brightest quiescent galaxies in the $z=0.83$ cluster MS 1054-03 from \cite{Holden2010}. The data and model spectra have been smoothed for display purposes only.}
\label{fig:keck_examplefit}
\end{figure}

\begin{figure*}[]
\centering
\includegraphics[width=2\columnwidth]{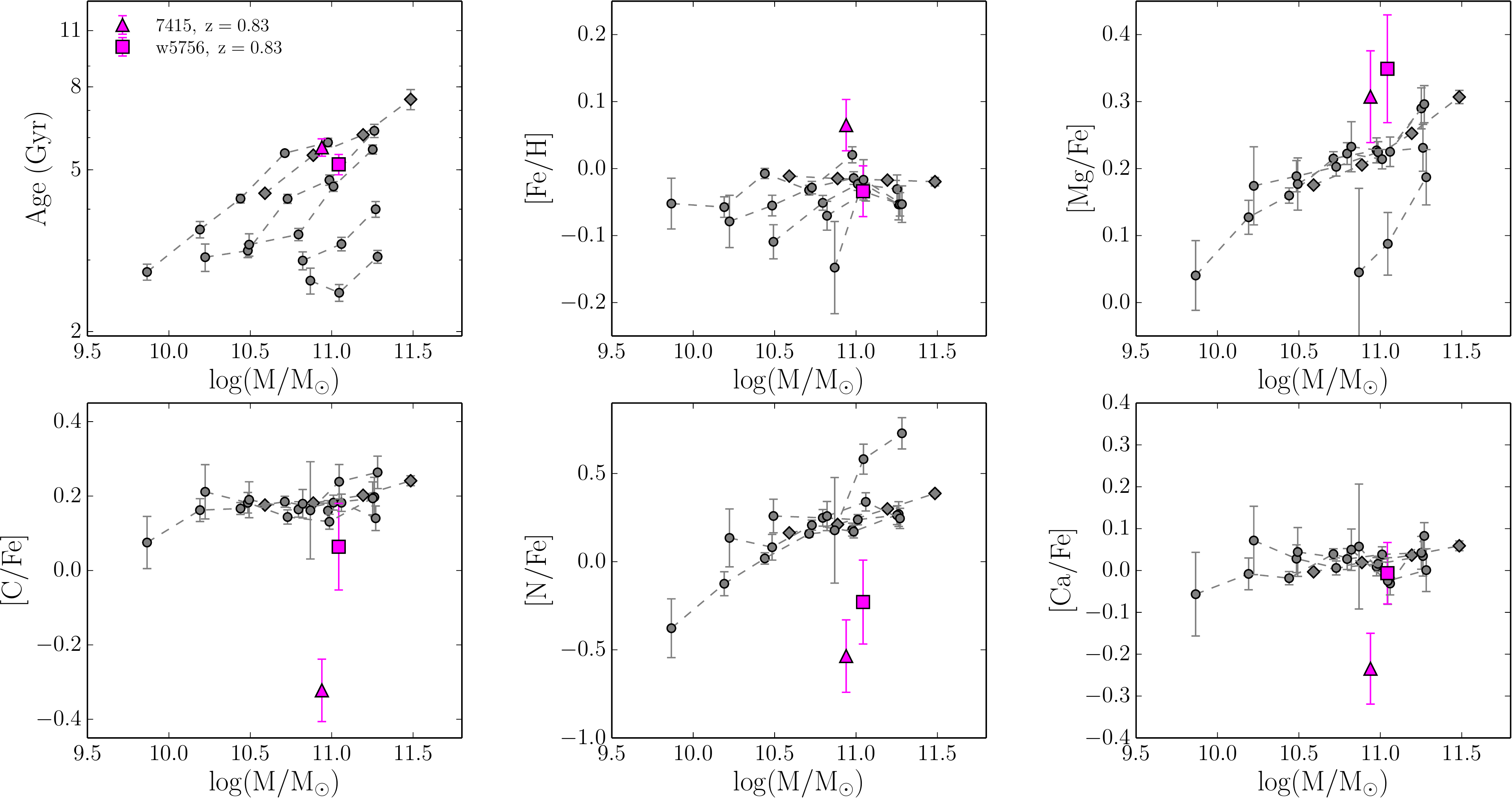}
\caption{Same as Figure~\ref{fig:ages_sdss_mass}, but now including the two $z=0.83$ cluster galaxies. The AGES and SDSS data are shown in gray symbols for comparison.}
\label{fig:ages_sdss_keck_mass}
\end{figure*}

\begin{figure*}
\centering
\includegraphics[width=1.92\columnwidth]{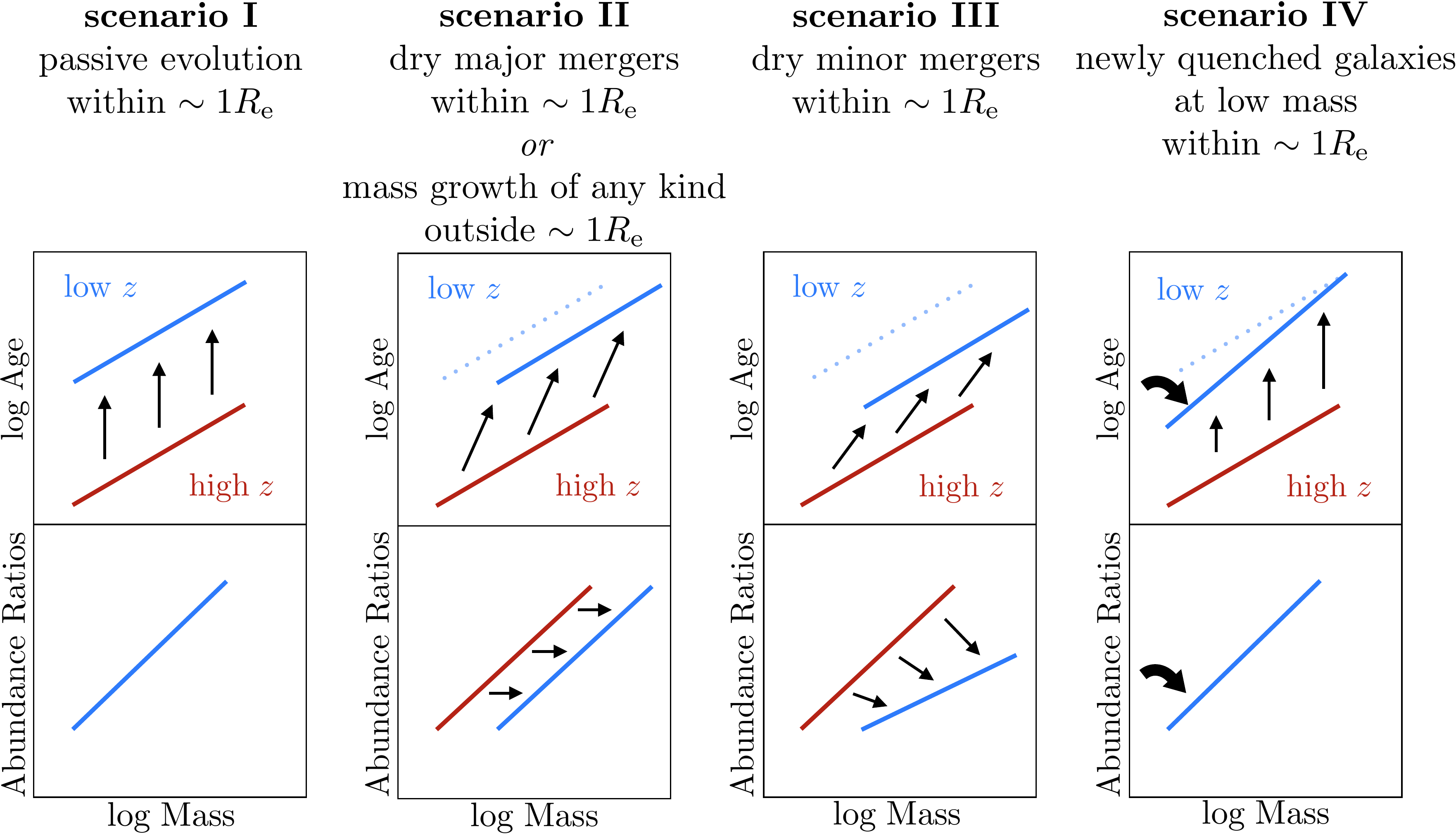}
\caption{Schematic diagrams for the time evolution of age and abundance ratios (e.g., [Mg/Fe]) of quiescent galaxies as a function of their stellar mass for four scenarios. Ages and abundance ratios are assumed to be measured within a spectroscopic fiber which has an extent of $\sim1~R_{\rm e}$, unless noted otherwise. The stellar mass spans $\approx1$~dex, the age spans $\approx0.7$~dex, and the abundance ratio spans $\approx0.3$~dex in all panels. At low masses, galaxies are assumed to have solar-scaled abundance ratios. {\it First panel}: passive evolution of isolated galaxies. The galaxies simply age over time, and the trend in abundance ratios remains unchanged. {\it Second panel}: dry major mergers (1:1) where a given galaxy doubles it mass, or mass growth of any kind occurring outside the fiber resulting in the doubling of total mass. For comparison, the low-redshift relation for passive evolution is shown as a dotted blue line. The major merger of two similar galaxies results in a more massive galaxy with stellar population properties that are unchanged. This simply moves the new galaxy to the right toward higher masses. This scenario is also equivalent to a passively evolving stellar population in the inner region accompanied by mass buildup in the outskirts. The accumulation of disrupted galaxies and/or in situ star formation at large radius leads to an increase in total stellar mass and shifts the galaxy to the right, but the spectroscopic data are insensitive to these outer regions and thus stellar population parameters remain unchanged. {\it Third panel}: dry minor mergers (1:10) within $\sim1~R_{\rm e}$ where a given galaxy doubles its mass. The accretion of a lower mass galaxy ``dilutes'' both age and abundances, shifting the new galaxy both downward and to the right from the original relation. For an equal fractional gain in mass, the changes introduced to both age and abundance ratios are thus larger for minor mergers compared to major mergers. {\it Fourth panel}: introduction of newly quenched low-mass galaxies to the passive population. New galaxies are assumed to have solar-scaled abundance ratios. The massive end evolves passively with the universe, but the ages at the low-mass end are lowered due to the presence of younger stellar populations. It should be noted that these scenarios are applicable only for $z\lesssim1$. At the highest redshifts, these relations should break down and the [$\alpha$/Fe] abundances are predicted to be super-solar at all stellar masses. This is because a quenched galaxy at very high redshift, regardless of its stellar mass, has to have formed its stars on short timescales to be consistent with the very young age of the universe.}
\label{fig:schematic}
\end{figure*}

However, [Mg/Fe] abundances for the two lowest mass bins in the highest redshift interval (blue points in top right panel of Figure~\ref{fig:ages_sdss_mass}) noticeably deviate from the main trend, albeit with large error bars. Mg abundance measurements rely on the Mg~$b$ feature at $\sim5175$~\AA, which corresponds to $8000$~--~8800~\AA{} in the observed frame for $0.55<z<0.7$. The Mg~$b$ feature at these high redshifts unfortunately falls in the spectral region affected by atmospheric absorption as well as sky emission features. In particular, there is an H$_2$O band spanning the wavelength interval $8000$~--~8400~\AA{} \citep{Palle2009}. Between $z=0.545$ and $z=0.627$, the Mg $b$ absorption feature is redshifted to wavelengths that overlap this atmospheric absorption feature. Since contamination affects continuum-normalization as well as the shape of the Mg~$b$ feature, we have carried out a test where we selected narrow redshift intervals that avoid the H$_2$O band, stacked the spectra, and remeasured the Mg abundances. The resulting Mg abundances were consistent with the original results and remained low. We have performed an additional test by stacking a subset of spectra with median S/N~$>5$~\AA{}$^{-1}$ per galaxy (the highest 50\% S/N in the redshift bin). The resulting Mg abundances were again consistent with the original results. As abundance ratios are generally expected to remain constant or decrease, but not increase, with time (see Figure~\ref{fig:schematic}), we remain skeptical regarding the Mg abundances at these high redshifts. A definitive resolution will require more and deeper spectra, and very careful attention to the sky subtraction and telluric corrections.

Similar to Mg, N and C also increase with mass, though the slope for C is modest and the scatter is larger compared to that of N. However, as in the case of [Fe/H], there is no redshift evolution in these trends. Although formally an $\alpha$ element, Ca tracks Fe and is close to solar. The interpretation is that a significant fraction of observed Ca originates from Type Ia SNe along with the iron-peak elements, rather than from massive stars \citep{Nomoto1984}. {\it In summary, there is no compelling evidence for redshift evolution in the abundances of quiescent galaxies at a fixed stellar mass}.

Figure~\ref{fig:logage_redshift} offers another way of examining the age evolution of quiescent galaxies by showing the evolution with redshift of galaxies at fixed stellar mass. Each set of colored symbols shows a single mass bin from the AGES sample for the three highest mass bins. Lower mass bins are not shown because they do not span a wide range in redshift. The age corresponds to the SSP-equivalent age of the representative galaxy in each mass--redshift bin. The black curves represent the evolution of light-weighted age from FSPS, v2.4 \citep{Conroy2009, Conroy2010} models with different star formation histories of the form ${\rm SFR}(t) \propto \exp(-t/\tau)$. These are solar-metallicity and dust-free models, and the light-weighted ages are measured at 5000~\aa{}. In the top panel, the tracks correspond to passive evolution following a burst ($\tau=0.1$~Gyr) at three different formation redshifts, $z_{\rm f}$. In the bottom panel, the models have the same $z_{\rm f}=3$ but different $\tau$. If $z_{\rm f}=2$ is assumed instead, the equivalent track corresponds to a value of $\tau$ that is smaller by 0.5 Gyr. The extended star formation histories associated with a large $\tau$ value are conceptually equivalent to different galaxies in the considered sample undergoing star formation with different starting and ending times, thus implying the inhomogeneous nature of the $z<1$ quiescent population. Regardless of the details of the adopted star formation histories, massive galaxies in our sample appear to be passively evolving for over half of cosmic time, though more data at $z\gtrsim1$ will be required to differentiate between the different models. More specifically, the age evolution of the highest-mass galaxies is strongly consistent with passive evolution while there is evidence for a shallower age evolution for the lower mass galaxies shown in this figure. This suggests that the age evolution of these lower mass galaxies is diluted, or slowed down, by the addition of newly-quenched, young galaxies. At even lower masses ($\lesssim10^{10.5}~\msun$), the data do not span a large enough range in redshift to make definitive statements. 

\subsection{Stellar Population Parameters of Quiescent Galaxies in a z=0.83 Galaxy Cluster from Keck DEIMOS Spectra}
\label{section:keck_results}
In addition to the stacked spectra of SDSS and AGES galaxies, we also fit spectra of two individual galaxies (7415 and w5756) in the cluster MS 1054-03 at $z=0.83$. Figure~\ref{fig:keck_examplefit} shows the data and the best-fit model. The continuum-normalized fluxes are shown in black while the models are plotted in red in the top panels. The model and data flux have been smoothed by a 3-point boxcar for display purposes only. The bottom panels show the percent fractional residuals between the best-fit models and data in black and the flux uncertainties as the gray shaded region. The region around 4200~\aa{} is masked out and not fit due to the location of the atmospheric \emph{A}-band in the observed frame, and the $\sim5100$~\aa{} region is also masked out due to the presence of strong sky OH emission lines. Overall, the quality of the fit is very good and the residuals are consistent with the flux uncertainties.

The resulting best-fit parameters are displayed in magenta symbols in Figure~\ref{fig:ages_sdss_keck_mass} along with the SDSS and AGES points shown in gray. The stellar masses of the galaxies come from combining the reported $r$-band magnitudes from \cite{Holden2010} and $M/L$ output from modeling the spectra. Both galaxies are older than expected from the AGES results but still consistent with the age of the universe at that epoch. At $z=0.83$, the universe is approximately 6.5~Gyr old, and the galaxies are measured to be approximately 5~Gyr in age. We note that previous studies of the $M/L$ ratios of massive quiescent galaxies did not find large differences between galaxies in low- and high-density environments \citep[e.g.][]{vanDokkum2007}. This may be caused by a difference in selection criteria: many studies of the $M/L$ ratios of galaxies in low-density environments selected objects by color and morphology rather than SFR \citep[see, e.g.,][]{vanderWel2005}. The abundances of these cluster galaxies are broadly consistent with the AGES and SDSS trends. However, [C/Fe] and [Ca/Fe] for 7415 and [N/Fe] for both objects are low, though this may be due to the influence of the \emph{A}-band which is partially polluting the CN feature.

These results serve as interesting and sobering test cases for modeling stellar populations of galaxies at progressively higher redshifts. Although our models work well at low S/N on idealized test cases (see Appendix~\ref{section:snr_test}), the reliable measurement of stellar population parameters also requires clean rest-frame optical spectra. As evidenced here and also in the previous section for the AGES spectra at $0.55<z<0.7$, obtaining high-quality spectra becomes very challenging at these redshifts due to the rest-frame optical becoming redshifted to the observed-frame NIR, where telluric absorption and sky emission features dominate. We therefore caution against over-interpreting the results in the Appendix when applied to data with strong systematic uncertainties due to sky subtraction and telluric features.

\section{DISCUSSION}
\label{section:discussion}
\subsection{The Assembly Histories of Quiescent Galaxies}
Our results indicate that for quiescent galaxies, there is negligible evolution in elemental abundances ($\lesssim0.1$~dex) at fixed stellar mass over roughly 7~Gyr, and the increase in their ages with cosmic time is consistent with passive evolution since $z=0.7$ (see Figures~\ref{fig:ages_sdss_mass} and \ref{fig:logage_redshift}). The stellar masses from SED-fitting nominally represent the total stellar mass. However, since our stellar population analysis is based on spectra obtained with a $1\farcs5$ fiber, the measured parameters are sensitive to stellar populations in the inner regions of galaxies directly probed by the fiber. There are two effects at play: the angular size of galaxies of fixed physical size increases with decreasing redshift and galaxies undergo intrinsic size growth with decreasing redshift. The implications of fiber aperture bias are investigated in Section~\ref{section:gradients}, where we demonstrate that this has a negligible effect on our conclusions. The conclusion from that section is that we are probing the evolution of the inner $\sim0.3\text{--}3~R_{\rm e}$ of massive quiescent galaxies.

We compare the main results shown in Figure~\ref{fig:ages_sdss_mass} with simple conceptual models in an attempt to interpret the trends in the context of the assembly histories of galaxies. We consider four scenarios, graphically represented in Figure~\ref{fig:schematic}. The age and abundance ratio (e.g., [Mg/Fe]) trends with mass are well-known in the local universe. In all four panels, we consider a high-redshift population of quiescent galaxies and trace their evolution over time using their age and abundances, rather than starting with the $z=0$ relation and extrapolating back in time. The stellar population parameters are assumed to be measured within a spectroscopic fiber which has an extent of $\sim1~R_{\rm e}$, unless noted otherwise. The stellar mass spans \mbox{$\approx1$~dex}, the age spans $\approx0.7$~dex, and the abundance ratio spans $\approx0.3$~dex in all panels. At low masses, galaxies are assumed to have solar-scaled abundance ratios.

It should be noted that these scenarios are applicable only for $z\lesssim1$. At the highest redshifts, these relations should break down and the [$\alpha$/Fe] abundances are predicted to be super-solar at all stellar masses. This is because a quenched galaxy at very high redshift, regardless of its stellar mass, has to have formed its stars on short timescales to be consistent with the very young age of the universe. 

The first scenario shows the expected evolution of a population of isolated, passively evolving quiescent galaxies. The galaxies simply age over time, and the trend in abundance ratios remains unchanged. For the next two panels, we assume that a given galaxy doubles its total stellar mass through one or more merger events. The second panel illustrates the evolution of quiescent galaxies undergoing dry (i.e., no star formation) major mergers (1:1) or mass growth of any kind occurring outside the fiber resulting in the doubling of total mass. For comparison, the low-redshift relation for passive evolution is shown as a dotted blue line. The major merger of two similar galaxies results in a more massive galaxy with stellar population properties that are unchanged. This simply moves the new galaxy to the right toward higher masses. This scenario is also equivalent to a passively evolving stellar population in the inner region accompanied by mass buildup in the outskirts. The accumulation of disrupted galaxies and/or in-situ star formation at large radius leads to an increase in total stellar mass and shifts the galaxy to the right, but the spectroscopic data are insensitive to these outer regions and thus stellar population parameters remain unchanged. The third panel shows the evolution of quiescent galaxies undergoing dry minor mergers (1:10) within $\sim1~R_{\rm e}$. The accretion of a lower mass galaxy ``dilutes'' both age and abundances, shifting the new galaxy both downward and to the right from the original relation. For an equal fractional gain in mass (i.e., a factor of two in this case), the changes introduced to both age and abundance ratios are thus more severe for minor mergers compared to major mergers. The fourth scenario is a passive evolution model that includes the addition of recently quenched low-mass galaxies, which are assumed to have solar-scaled abundance ratios. The massive end evolves passively with the universe, but the ages at the low-mass end are lowered due to the presence of recently formed stellar populations. These illustrated relations are only meant to represent a simple schematic picture, since the exact relations and relative slopes depend on the details of the merger history as well as the age and abundance distributions of the new quiescent galaxies. Nevertheless, these simple models highlight the power of the measured abundances combined with the age to probe the evolution of quiescent galaxies. All scenarios produce subtle changes in age, which are difficult to reliably detect owing to the systematic uncertainties affecting stellar age measurements. However, in the abundance-mass space, the changes are more pronounced and qualitatively different, thereby enabling greater distinction between the different scenarios.

Our main result is that at fixed stellar mass, the abundance ratios at difference redshifts vary by less than 0.1~dex for most cases. This conclusion is most robust for the massive galaxies ($>10^{10.5}~\msun$) because there are more data spanning a large redshift range at higher masses. The lack of evolution with cosmic time in the mass-abundance ratio correlation leads us to favor scenario I. Taken at face value, these results favor a scenario in which the inner $\sim0.3\text{--}3~R_{\rm e}$ of massive quiescent galaxies have been passively evolving for the past $\sim7$~Gyr. As we demonstrate at the end of this section, the results are also consistent with modest mass growth between $z=0.7$ and $z=0.1$ in the outskirts (scenario II).

Due to the absence of lower mass galaxies at high redshifts in our sample, we cannot draw definitive conclusions about the evolution of these objects from the data. However, the young ages of these galaxies combined with their roughly solar-scaled abundances indicate that our results are consistent with the addition of new, low-mass quiescent galaxies with solar-scaled abundance ratios (see scenario IV in Figure~\ref{fig:schematic}). The addition of low-mass quiescent galaxies over time is also favored by simulations \citep[e.g.,][]{Cen2014} as well as independent data, such as evolution in the luminosity and mass functions, which indicate that the number of red galaxies has doubled since $z=1$ \citep[e.g.,][]{Faber2007, Pozzetti2010, Moustakas2013}.

As demonstrated in Figure~\ref{fig:logage_redshift}, the derived SSP-equivalent ages are considerably younger than the age of the universe at all epochs, suggesting an \emph{equivalent} single-burst star formation epoch of $z_{\rm f}\lesssim1.5$. We stress that $z_{\rm f}$ is a representative parameter, and not suggestive of the {\it actual} formation epoch. The real SFH is likely more complex and extended in time, but it is a common point of comparison to represent complex SFHs with an SSP-equivalent, or effective single-burst, epoch \citep[e.g.,][]{Treu2005, vanDokkum2007}. In other words, this result should not be interpreted as the galaxies in each mass--redshift bin having uniformly formed their stars in a single burst at $z_{\rm f}\lesssim1.5$. Instead, the low value of $z_{\rm f}$ implies that our results are inconsistent with all of the stars in the galaxies in our sample having formed at very high redshifts \cite[see also the discussion in ][]{Schiavon2006}. The addition of newly quenched galaxies at $z_{\rm f}\gtrsim1$ \citep{vanDokkum2001} naturally explains the young ages of galaxies in our sample, and is supported by both simulations and observations \citep[e.g.,][]{Whitaker2010, ElicheMoral2010, Prieto2013, Cen2014, Marchesini2014}. In order to be consistent with the apparent passive evolution since $z=0.7$, young quiescent galaxies cannot be entering the sample in large numbers at these late times at the highest masses.

The inhomogeneous nature of the $z<0.7$ quiescent population suggests that there may be variations in the ages of galaxies within a given mass--redshift bin (see also \citealt{Whitaker2010}). We have tried performing unweighted stacking to ensure that a few bright, young galaxies, which are given more weight during the stacking procedure due to their smaller flux errors, are not driving the ages to low values. The resulting best-fit ages as well as the abundances agree with results from weighted stacking to within 1$\sigma$ errors, thereby demonstrating that this is a negligible effect. In addition, there is no correlation between S/N and $U-V$ color of individual galaxies within each mass--redshift bin, indicating that there is no obvious bias against the reddest galaxies, i.e., the reddest galaxies are equally likely to have large weights in the stacks as the bluest galaxies in the sample. As discussed in Appendix~\ref{section:stacking_test}, the best-fit parameter measured from the stacked spectrum and the weighted average of best-fit parameters from fitting individual spectra agree to within $0.05$~dex. Moreover, the unweighted average of the best-fit ages resulting from fitting individual galaxies agrees with the weighted average age to within 1~Gyr. Thus we conclude that these low ages are representative of the galaxies in the sample, though it should be noted that these are still SSP-equivalent ages.

Our results are also consistent with the conclusions from size evolution studies \citep{Daddi2005, Trujillo2006, vanDokkum2008, vanderWel2008, Cimatti2008, Bezanson2009, Damjanov2009, Williams2010, Cassata2010, vanDokkum2010, LopezSanjuan2012, McLure2013, Belli2013}. For example, \cite{vanDokkum2010} found that stellar mass in the central regions (inner 5~kpc) has remained roughly constant with redshift, while the outer regions (out to $\sim75$~kpc) of massive quiescent galaxies have been gradually building up over the last 10~Gyr. In other words, stellar populations have been passively evolving in the centers of massive quiescent galaxies, undisturbed by bursts of star formation or merger activities, while the outskirts have been evolving overtime. Mass and size growth is thought to occur mostly via minor mergers, although in-situ star formation may contribute $\sim20\%$ to the total mass buildup \citep{vanDokkum2010}.

As previously discussed, stellar masses quoted in this work nominally represent the total stellar mass. As noted above, the size growth of massive galaxies suggests that they are growing in total mass as well. However, the amount of mass growth from $z=0.7$ to $z=0.1$ is not dramatic, amounting to $\lesssim30\%$ or $\lesssim0.15$ dex \citep{vanDokkum2010}, which is comparable to the uncertainties in our stellar mass estimates. Moreover, since the adopted mass bins for creating stacked spectra are $\gtrsim0.3$ dex wide, the effects of modest mass growth (e.g., 0.15~dex) will not be discernible in the present work. In other words, our main results taken at face value favor a passively evolving stellar population since $z=0.7$, but they are also consistent with modest mass growth in the outskirts over time.

\subsection{Comparisons to Previous Work}
Our results broadly agree with previous work on stellar populations in quiescent galaxies at low redshift \citep[e.g.,][]{Trager2000b, Saglia2002, Cenarro2003, Eisenstein2003, Thomas2003, Worthey2004, Thomas2005, Sanchez2006b, Graves2007, Schiavon2007, Graves2008, Matkovic2009, Smith2009, Zhu2010, Johansson2012, Worthey2014, Conroy2014}. The primary conclusions from low-redshift studies, including our own, is that more massive galaxies are older, more metal-rich, and have enhanced abundance ratios in the elements Mg, C, and N compared to lower mass galaxies.

Owing to the demanding S/N requirements, there have been relatively few studies\footnote{During the refereeing process, we became aware of \cite{Gallazzi2014}, who analyzed optical spectra of $\sim70$ star-forming and quiescent galaxies in the redshift range $0.65\geq z \geq 0.75$ and obtained relations between light-weighted stellar age, stellar mass, and stellar metallicity for both the total galaxy population and for star-forming and quiescent galaxies separately. The authors concluded that both the metallicity and age evolutions of the quiescent galaxies at $z=0.7$ are consistent with passive evolution, in excellent agreement with our results.} of stellar abundances patterns at $z\gtrsim0.1$. \cite{Kelson2006} examined 19 cluster elliptical and lenticular galaxies at $z=0.33$ and measured the age and abundance ratios using eight blue Lick/IDS indices. Interestingly, while they found that total metallicity and [N/Fe] are tightly correlated with velocity dispersion, they did not find significant variations in age, [$\alpha$/Fe], nor [C/Fe] with velocity dispersion.

\cite{Sanchez2009} analyzed stacked spectra of 215 red-sequence galaxies in cluster and group environments from $z\sim0.75$ to 0.45 using Lick/IDS indices. The authors confirmed that massive galaxies have $Z\sim \rm Z_{\odot}$ and are enhanced in $\alpha$-elements. Additionally age variation was found to be consistent with passive evolution, all in agreement with the present work. They concluded that massive galaxies formed their stars at high redshift and passively evolved with time, while lower mass galaxies experienced longer star formation episodes and thus have constant luminosity-weighted ages over the redshift range considered. Also in qualitative agreement with our results, they demonstrated that the total metallicity and [$\alpha$/Fe] as a function of velocity dispersion (or mass) are constant with time.

\cite{Schiavon2006} went beyond the cluster environment and pushed to even higher redshifts, analyzing Keck DEIMOS spectra of $0.7\leq z \leq 0.9$ red field galaxies. Their stellar population synthesis modeling showed that $z\sim0.9$ galaxies have mean light-weighted ages of only 1~Gyr. The authors concluded that either individual galaxies are experiencing low-level star formation (i.e., ``frosting") or galaxies with younger stars are continually being added to the quiescent population. Interestingly, the time elapsed between $z\sim0.9$ and $z\sim0.7$ is roughly 1 Gyr, and we estimate the youngest $z\sim0.7$ AGES galaxies to be approximately 2--3 Gyr in age. It is thus plausible that the $z\sim0.9$ quiescent galaxies from \cite{Schiavon2006} are progenitors of the $z\sim0.7$ galaxies in our sample. 

Recently, \cite{Jorgensen2013} analyzed the stellar populations of quiescent galaxies in three galaxy clusters at $z=0.54$, 0.83, and 0.89, and inferred that while the evolution in the fundamental plane is consistent with passive evolution, the variations in total metallicity and [$\alpha$/Fe] with redshift appear to be inconsistent with a passive evolution scenario in the redshift interval considered. Their velocity dispersion-line indices scaling relations indicate that the blue metal lines are stronger than expected for passive evolution. 

At least some of the discrepancies between conclusions of previous work and our favored interpretation can be attributed to differences in the analysis (e.g., different stellar population synthesis models) and sample selection. The present work selects quiescent galaxies based on sSFR estimated from SED-fitting, but other options for quiescence selection include various combinations of morphological, emission line, S/N, and/or color cuts. In contrast to our quiescent sample which was selected to be mass-complete, most previous high-redshift studies specifically targeted group and cluster environments (of the four discussed above, only \citealt{Schiavon2006} examined field galaxies) due to the efficiency of obtaining many simultaneous spectra with current generation multi-object spectrographs. Typically a few groups or clusters were analyzed at a time, rendering the task of connecting the results to the global quiescent galaxy population difficult as the results can be affected by small-number statistics. These factors make a straightforward and detailed comparison between our results and previous work very challenging.

The role of environment on stellar population properties at fixed stellar mass is a subject of ongoing debate in the literature \citep{Sanchez2003, Eisenstein2003, Thomas2005, Sanchez2006b, Trager2008, Toloba2009, Zhu2010, Thomas2010, Johansson2012}. In the present work, we have found that the Keck DEIMOS sample consisting of two bright quiescent galaxies in a $z=0.83$ cluster are older compared to the average quiescent galaxy at $z=0.7$, but their ages are consistent with the age of the universe at $z=0.83$. The results hint at the impact of environment on stellar populations of galaxies, but larger samples will be required to verify these tantalizing trends.

\subsection{Radial Gradients and Fiber Size}
\label{section:gradients}
An important effect that needs to be considered is the fraction of the galaxy observed as a function of both the fiber size and the redshift of the galaxy. The Hectospec instrument on the MMT has a fiber that subtends $1\farcs5$ on the sky, whereas the SDSS spectroscopic fiber has a diameter of $3\farcs$ For reference, $1\farcs5$ corresponds to 2.8~kpc and 10.7~kpc at $z=0.1$ and 0.7, respectively. There are two effects that need to be considered. First, the angular extent of a galaxy of fixed physical size is larger at smaller redshift. Second, quiescent galaxies are believed to have been growing in both size and mass over the last $\sim$10 Gyr \citep[e.g.,][]{Naab2009, vanDokkum2010}. This implies that we are sampling different fractions of the galaxy at different redshifts, both due to the varying angular diameter distance and to the intrinsic growth of the galaxy with redshift. In the presence of radial gradients, this could introduce a bias (with respect to the global properties of the galaxy) in the inferred stellar population parameters. 

Spatially resolved observations of galaxies at low redshift have enabled radial gradient measurements of various stellar population parameters \citep[e.g.,][]{Gorgas1990, Sanchez2007, Baes2007, Brough2007, Rawle2010, Spolaor2010, Loubser2012, LaBarbera2012, Greene2013}. \cite{Greene2013} measured the radial gradients of a sample of 33 massive quiescent galaxies and found a modest negative gradient in [Fe/H] and a strong radial decline in [C/Fe], but almost constant [Mg/Fe], [N/Fe], and [Ca/Fe] out to 2.5~$R_{\rm e}$. Additionally, the age gradient was measured to be very weak or nonexistent. Similarly, \cite{Spolaor2010} inferred almost no radial gradient in age and [$\alpha$/Fe] but a weak negative gradient in [Z/H] from a sample of 37 massive quiescent galaxies. From these local studies it thus appears unlikely that our age and [$\alpha$/Fe] measurements are biased by the sampling radius of the fiber, but we cannot immediately rule out the possibility that our [Fe/H] and [C/Fe] measurements are affected by radial gradients.

\begin{figure}
\centering
	\subfigure{
	\includegraphics[width=0.9\columnwidth]{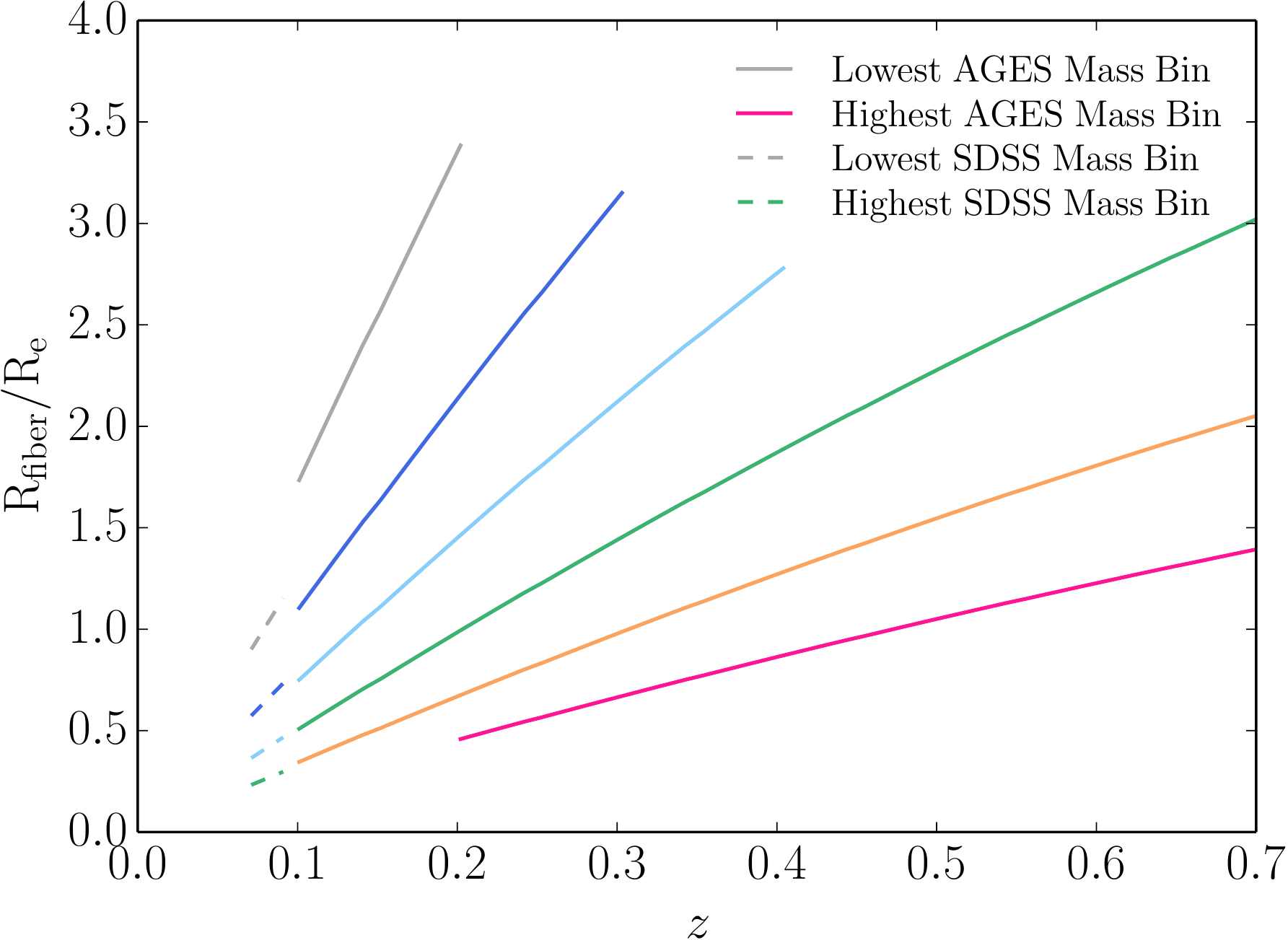}
	}
	\subfigure{
	\includegraphics[width=0.9\columnwidth]{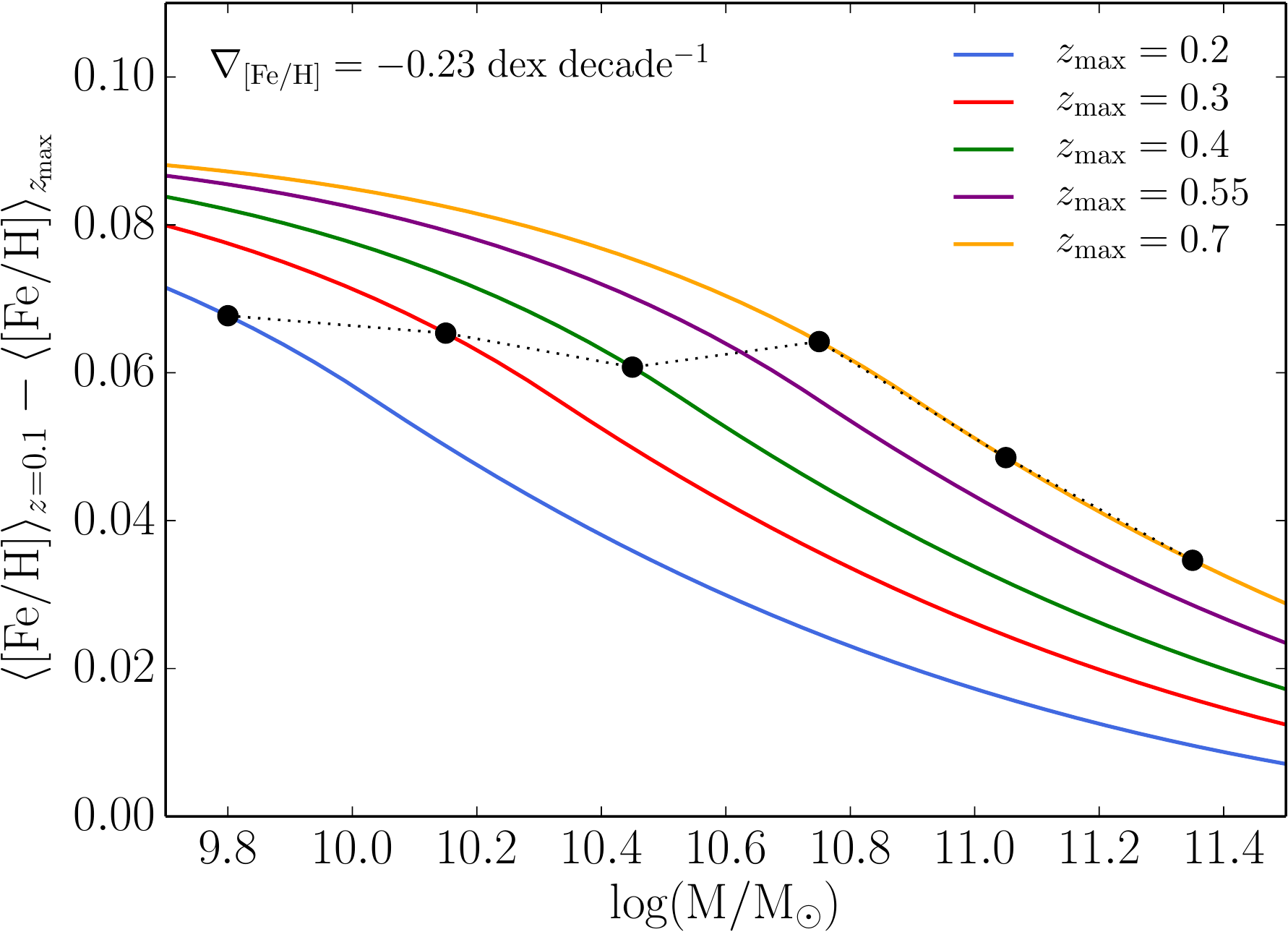}
	}
\caption{{\it Top panel:} the ratio of fiber radius to the effective radius as a function of redshift. The effective radius is computed by assuming a mass-size relation at $z\sim0$ from SDSS data \citep{Shen2003} and applying a redshift evolution scaling $R_{\rm e} \propto (1+z)^{-1}$ \citep[e.g.,][]{Williams2010}. Each curve spans the entire redshift range sampled by the data in each mass bin. The dashed and solid lines correspond to SDSS and AGES data, respectively. The mass bins increase from top to bottom (gray to pink). The mid-bin values for each mass interval are $\log (M/\msun) = 9.80$, 10.15, 10.45, 10.75, 11.05, and 11.35 for AGES, and $\log (M/\msun) = 10.58$, 10.93, 11.28, and 11.63 for SDSS. {\it Bottom panel:} the differences in inferred [Fe/H] between $z=0.1$ and five different values of $z$ ranging from 0.2 to 0.7, shown in different colored curves for a metallicity gradient of $\nabla_{\rm [Fe/H]} \equiv{\rm \Delta [Fe/H]}/\Delta \log {(R/R_{\rm e})}=-0.23$~dex~decade$^{-1}$. The black circles represent the maximum aperture bias between the lowest and highest redshift bins within each AGES stellar mass bin.}
\label{fig:gradients}
\end{figure}

We carry out simple simulations to quantitatively assess this possibility. We investigate [Fe/H] only, but similar conclusions can be drawn for [C/Fe]. We adopt a metallicity gradient from \cite{Spolaor2010}, corresponding roughly to $\nabla_{\rm [Fe/H]} \equiv{\rm \Delta [Fe/H]}/\Delta \log {(R/R_{\rm e})}=-0.23$~dex~decade$^{-1}$ over the range $0.5~R_{\rm e}$--$2.5~R_{\rm e}$, assuming that [Z/H]~$\propto$~[Fe/H]. The metallicity profile ${\rm [Fe/H]}(r)$ is convolved with an $n=4$ Sersic profile $I(r)$ to obtain a light-weighted average metallicity within a given radius, $\langle {\rm [Fe/H]}(r) \rangle$. The effects of seeing ($\approx1\farcs5$ for AGES, which is comparable to the diameter of the fiber) is ignored because while light is scattered out of the fiber, light from larger radius is also scattered inward. Thus light captured by the fiber is actually sampled from a larger radius, on average, thereby effectively reducing the magnitude of aperture bias. Here we define $f_{\rm origin}$, the fraction of light originating from outside the fiber radius that is scattered into the fiber, i.e., $f_{\rm origin}=0$ in the absence of seeing. In the high-$z$ limit where the fiber size and seeing are larger than the effective radius, suppose by a fiducial factor of three, $f_{\rm origin}~\approx0.1$. In the low-$z$ limit where the effective radius is roughly three times larger than both the fiber and the seeing, $f_{\rm origin} \approx0.3$, which is non-negligible. In the calculations below we do not include the effects of seeing because it provides the worst-case scenario for bias effects.

The bias introduced as a result of observing within an aperture of radius $r$ is calculated following \cite{Moustakas2011}:

\begin{equation}
\Delta \langle {\rm [Fe/H]}(r) \rangle = \langle {\rm [Fe/H]}(r) \rangle - \langle {\rm [Fe/H]}_{\infty} \rangle \;,
\end{equation}
where

\begin{equation}
\langle {\rm [Fe/H]}(r) \rangle = \frac{\int_{0}^{r}{\rm [Fe/H]}(r^\prime)I(r^\prime)r^\prime dr^\prime}{\int_{0}^{r} I(r^\prime)r^\prime dr^\prime}\;,
\end{equation}
and $\langle {\rm [Fe/H]}_{\infty} \rangle$ is the same quantity integrated out to $\infty$. For $\nabla_{\rm [Fe/H]} = -0.23$~dex~decade$^{-1}$, the bias introduced is relatively modest, corresponding to $\approx 0.08$~dex and $\approx 0.05$~dex when probing $0.5~R_{\rm e}$ and $2~R_{\rm e}$, respectively. When the gradient is twice as strong, akin to that found by \cite{Greene2013}, the resulting bias is roughly doubled.

The goal of this section is to investigate how much apparent evolution in [Fe/H] in our sample could be due to the effects of aperture bias. We estimate the average sizes of our quiescent galaxies as a function of redshift (direct size estimates are unfortunately not available for individual galaxies) to determine the fraction of the galaxy probed by the spectroscopic fiber as a function of redshift. To do this we assume a mass-size relation at $z\sim0$ from SDSS data \citep{Shen2003} and apply a redshift evolution scaling $R_{\rm e} \propto (1+z)^{-1}$ \citep[e.g.,][]{Williams2010}. The top panel of Figure~\ref{fig:gradients} shows the ratio of fiber radius to the effective radius as a function of redshift. Next, we compute the differences in inferred [Fe/H] between $z=0.1$ and five different values of $z$ ranging from 0.2 to 0.7 purely due to the sampled fraction of the galaxy evolving with redshift, shown as different colored curves in the bottom panel of Figure~\ref{fig:gradients}. The black circles represent the maximum aperture bias between the lowest and highest redshift bins within each AGES stellar mass bin. At lower masses, an apparent evolution of $\approx0.06$~dex is induced due to the evolving aperture bias, implying that for an intrinsically unevolving population, we would observe [Fe/H] to increase by $\approx0.06$~dex with time in the redshift range probed (e.g., from $z=0.4$ to $z=0.1$ for the third lowest mass bin). Apparent evolution is the strongest for low-mass galaxies, and the effect becomes weaker with increasing mass.

If radial gradients in [Fe/H] are present in our sample at the level we have assumed for this test, then this apparent evolution can masquerade as a true intrinsic evolution in stellar population parameters purely as a consequence of probing different fractions of the galaxy. However this is a small effect, amounting to at most $\lesssim0.06$~dex in our sample (or $\lesssim0.1$~dex in the more extreme case where the gradient is twice as strong for our sample of quiescent galaxies). Moreover, if we account for the effects of seeing, the magnitude of the bias would be even smaller. In comparison, the observed variation in the stellar population parameters as a function of redshift is $\lesssim0.1$~dex (see Figure~\ref{fig:ages_sdss_mass}). Thus we conclude that aperture bias has a minor effect on the interpretation of our results.

\section{Summary}
\label{section:summary}
In this paper we measured SSP-equivalent ages and detailed abundance patterns of quiescent galaxies using stacked SDSS and AGES spectra and individual spectra of two brightest $z=0.83$ cluster galaxies from \cite{Holden2010}. The main sample spans a redshift interval of $0.1<z<0.7$ and a stellar mass range from $10^{9.6}$ to $10^{11.8}~\msun$. We selected quiescent galaxies based on their star formation rates estimated from SED-fitting. The AGES sample of quiescent galaxies were divided into five redshift intervals each spanning roughly 1~Gyr in cosmic time and further divided into mass bins. The mass bins were chosen such that the sample was complete in stellar mass at each redshift. The stacked spectra were fit using a full spectrum modeling MCMC code developed by \cite{Conroy2012}. We also carried out a variety of systematic tests to examine the robustness of the modeling code. We now summarize our main results.

\begin{enumerate}
\item We confirm earlier results of stellar population modeling of quiescent galaxies at low redshift. Massive galaxies harbor old stellar populations with roughly solar [Fe/H] abundances and enhanced [Mg/Fe], [C/Fe], and [N/Fe] abundance ratios. Adopting [Mg/Fe] as a star-formation timescale indicator, massive galaxies seem to form their stars on shorter timescales compared to lower mass galaxies.

\item There is negligible evolution in the abundances of Fe, Mg, C, N, and Ca at fixed stellar mass over roughly 7~Gyr of cosmic time, and the evolution of the stellar ages of massive galaxies is consistent with passive evolution since $z=0.7$. The 0.1~dex or smaller variation in abundance ratios (e.g., [Fe/H], [Mg/Fe], [Ca/Fe]) as a function of stellar mass from $z=0.1$ to $z=0.7$ puts a stringent constraint on the assembly histories of galaxies over the redshift interval considered. Our results support the passive evolution of the inner $\sim0.3\text{--}3\;R_{\rm e}$ of massive quiescent galaxies ($M>10^{10.5}~\msun$) over the last $\sim7$ Gyr. At lower masses our results are also consistent with the addition of younger, newly quenched galaxies over time.

\item The derived SSP-equivalent ages are considerably younger than the age of the universe at all epochs, consistent with an \emph{equivalent} single-burst star formation epoch of $z_{\rm f}\lesssim1.5$. The addition of newly quenched galaxies at $z_{\rm f}\gtrsim1.5$ naturally explains the young ages of galaxies in our sample. These young stellar population ages coupled with the existence of massive quiescent galaxies at $z>1$ indicate the inhomogeneous nature of the $z\lesssim0.7$ quiescent population. In order to be consistent with the apparent passive evolution since $z=0.7$, young quiescent galaxies cannot be entering the sample at these late times at the highest masses.

\item There is tentative evidence that galaxies in cluster environments are older than galaxies inhabiting low-density environments. We hesitate to draw any strong conclusions, however, due to the small sample size as well as the possibility of contamination from atmospheric absorption and sky emission features imparting systematic effects on our spectral modeling results. This stresses the importance of obtaining deeper spectra with careful attention to telluric corrections and sky subtractions in addition to the development of tools capable of modeling low S/N spectra of these high-redshift galaxies.

\item The full spectrum fitting approach allows reliable and accurate abundance measurements, including age, Fe, Mg, C, N, and Ca, down to low S/N. The ability to accurately measure detailed abundance patterns in low S/N spectra with reliable uncertainty estimates opens the possibility to engage in detailed stellar population analysis at high redshift and in the low surface brightness outskirts of nearby galaxies.
\end{enumerate}

\acknowledgments{
We thank the referee for a prompt and constructive report. We also thank Tomer Tal for useful discussions and assistance with investigating the effects of radial gradients and Dan Foreman-Mackey for providing a custom Fortran implementation of {\tt emcee}. JC acknowledges support from the National Science Foundation Graduate Research Fellowship Program. CC acknowledges support from NASA grant NNX13AI46G, NSF grant AST-1313280, and the Alfred P. Sloan and Packard Foundations. MJIB acknowledges financial support from the Australian Research Council (FT100100280) and the Monash Research Accelerator Program (MRA).

This research has made use of observations obtained at the MMT Observatory, a joint facility of the University of Arizona and the Smithsonian Institution, and at the W.M. Keck Observatory, which is operated as a scientific partnership among the California Institute of Technology, the University of California and the National Aeronautics and Space Administration. W.M. Keck Observatory was made possible by the generous financial support of the W.M. Keck Foundation. Funding for the SDSS and SDSS-II has been provided by the Alfred P. Sloan Foundation, the Participating Institutions, the National Science Foundation, the U.S. Department of Energy, the National Aeronautics and Space Administration, the Japanese Monbukagakusho, the Max Planck Society, and the Higher Education Funding Council for England. The SDSS Web site is http://www.sdss.org/. The SDSS is managed by the Astrophysical Research Consortium for the Participating Institutions. The Participating Institutions are the American Museum of Natural History, Astrophysical Institute Potsdam, University of Basel, University of Cambridge, Case Western Reserve University, University of Chicago, Drexel University, Fermilab, the Institute for Advanced Study, the Japan Participation Group, Johns Hopkins University, the Joint Institute for Nuclear Astrophysics, the Kavli Institute for Particle Astrophysics and Cosmology, the Korean Scientist Group, the Chinese Academy of Sciences (LAMOST), Los Alamos National Laboratory, the Max-Planck-Institute for Astronomy (MPIA), the Max-Planck-Institute for Astrophysics (MPA), New Mexico State University, Ohio State University, University of Pittsburgh, University of Portsmouth, Princeton University, the United States Naval Observatory, and the University of Washington.
}

\begin{appendix}
\section{Quantifying Systematic Uncertainties}
\label{section:systematics}

\begin{figure*}
\centering
\includegraphics[width=1.94\columnwidth]{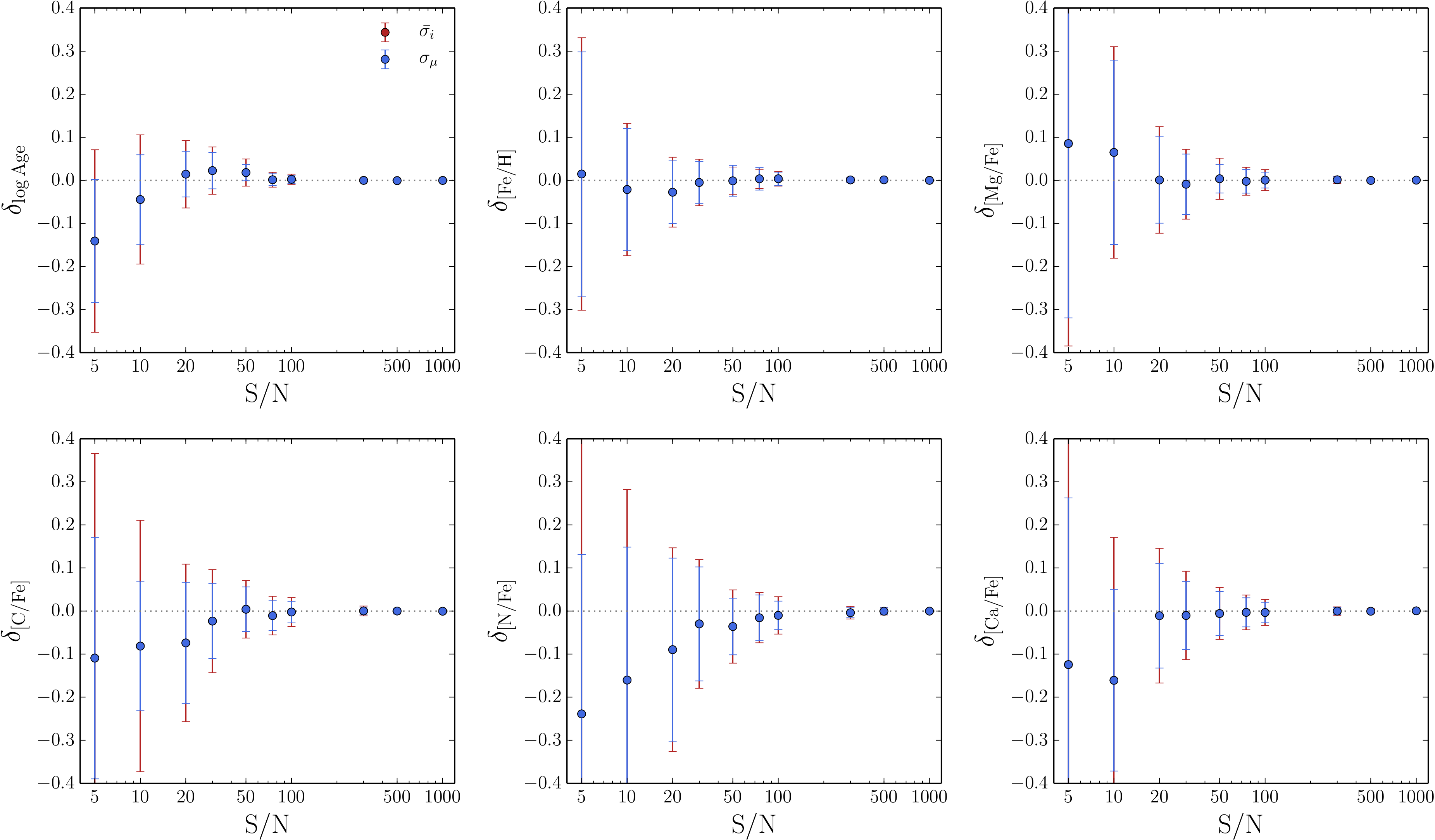}
\caption{Difference between parameters measured from artificially degraded spectra and the original high-quality SDSS stacked spectrum. We construct 50 realizations at each S/N. The different colored symbols represent two independent error estimates, where the red is the average of the 50 errors measured by the fitting code and the blue corresponds to the 1$\sigma$ scatter of the 50 measured parameters. Age and [Fe/H] are accurately recovered without significant systematic offsets down to S/N~$\approx10~\text{\aa{}}^{-1}$. [Mg/Fe] and [Ca/Fe], on the other hand, require S/N~$\approx20~\text{\aa{}}^{-1}$, and [C/Fe] and [N/Fe] demand S/N~$\approx30~\text{\aa{}}^{-1}$.}
\label{fig:snr_test}
\end{figure*}

\begin{figure*}
\centering
\includegraphics[width=1.94\columnwidth]{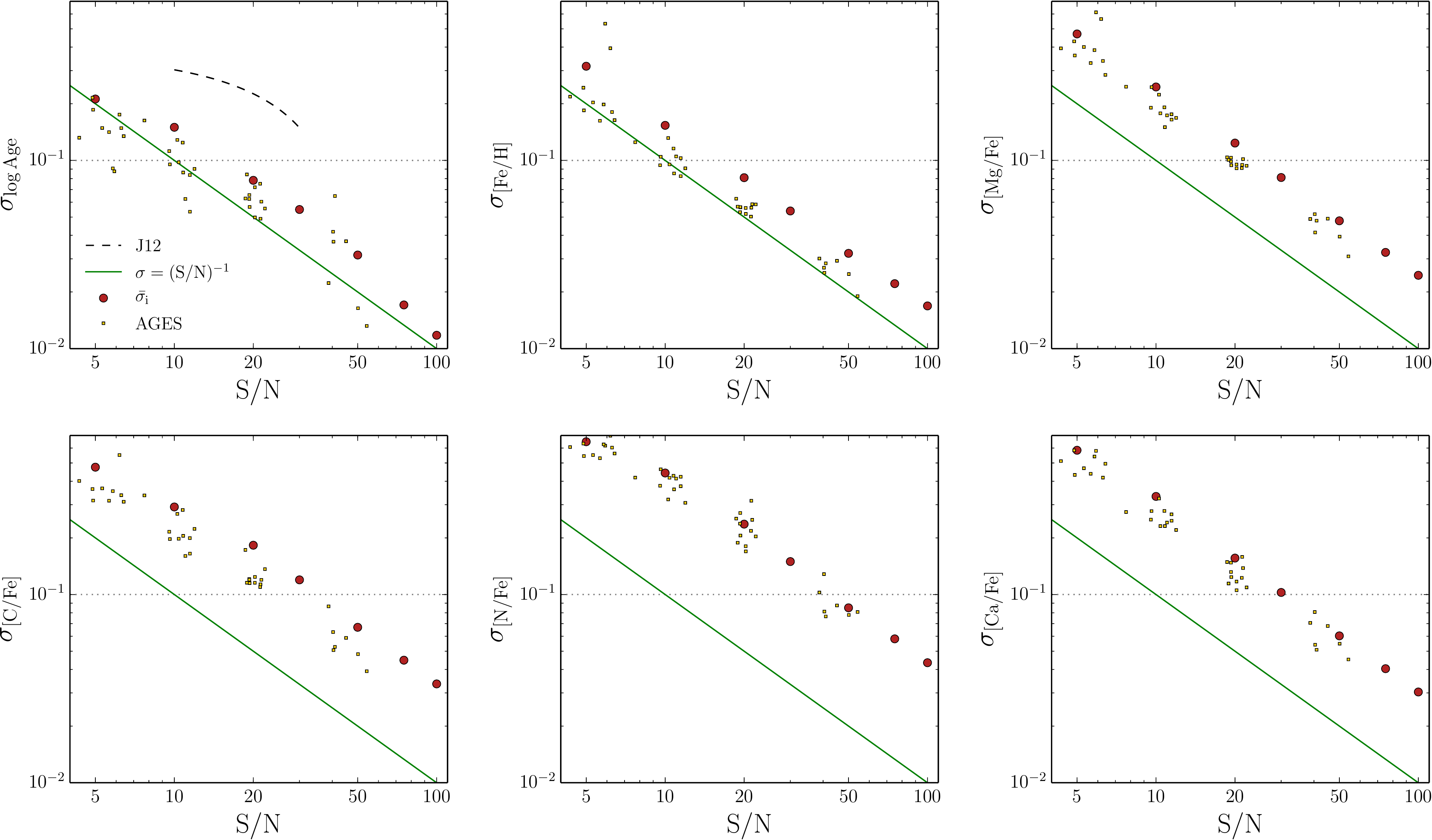}
\caption{Uncertainty on the parameters as a function of S/N per Angstrom. The solid green line is the theoretical expectation, and the dashed line labeled J12 is the S/N relation from \cite{Johansson2012} derived from fitting Lick indices to the spectra. The yellow points are errors on the best-fit parameters from individual AGES spectra, and the red symbols are the average of the 50 errors measured by the fitting code (the same red error bars from Figure~\ref{fig:snr_test}). Comparison in this figure between the simulated Gaussian noise result and the real AGES data shows that our assumption of ideal, uncorrelated noise for the former is a reasonable approximation. In addition, the errors are well-behaved and scale as expected. Age and [Fe/H] are reliably measured to $\approx0.1$~dex errors at S/N~$\approx10~\text{\aa{}}^{-1}$, while [Mg/Fe] and [Ca/Fe] are reliably measured to $\approx0.1$~dex errors at S/N~$\approx20~\text{\aa{}}^{-1}$. [C/Fe] and [N/Fe], on the other hand, require S/N~$\approx30\text{--}50~\text{\aa{}}^{-1}$ for $\approx0.1$~dex errors. }
\label{fig:errorbar_sn}
\end{figure*}

\begin{figure*}
\centering
\includegraphics[width=1.94\columnwidth]{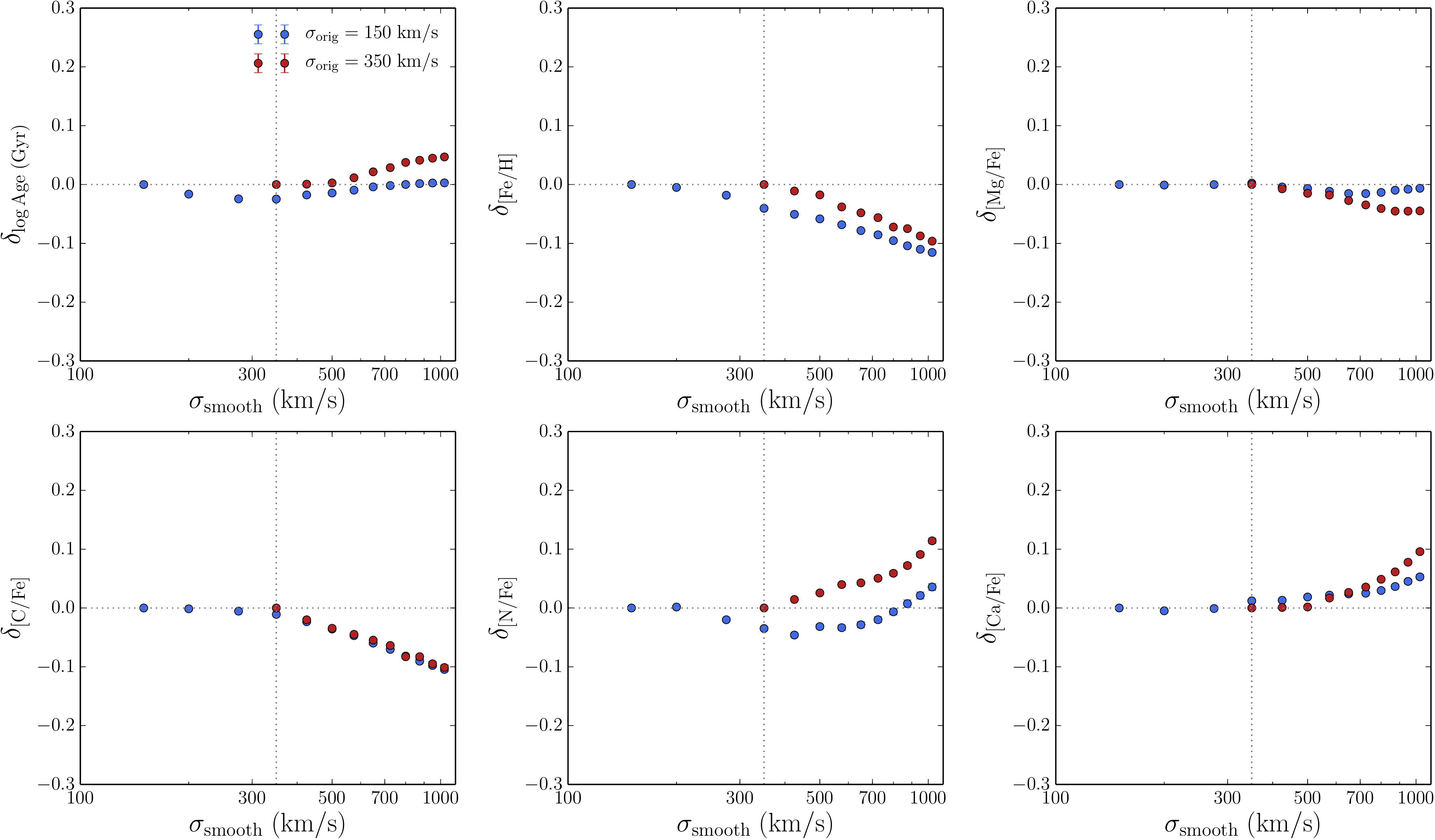}
\caption{Difference between parameters measured from artificially broadened spectra and the original unsmoothed SDSS stacked spectra. The vertical dotted line denotes $\sigma=350$~\kms, the effective velocity dispersion to which we smooth all of our science spectra. When $\sigma<150$ \kms{} spectra are smoothed to 350~\kms{}, there is almost no bias for [Mg/Fe], [C/Fe], and [Ca/Fe]. On the other hand, only a modest systematic offset of $\lesssim0.05$~dex is introduced at 350~\kms{} for age, [Fe/H], and [N/Fe]. In addition, there are only modest offsets ($\lesssim0.1$~dex) when the spectra are smoothed to $\sigma=1000$~\kms{}.}
\label{fig:smoothing_test}
\end{figure*}

\begin{figure*}
\centering
\includegraphics[width=1.94\columnwidth]{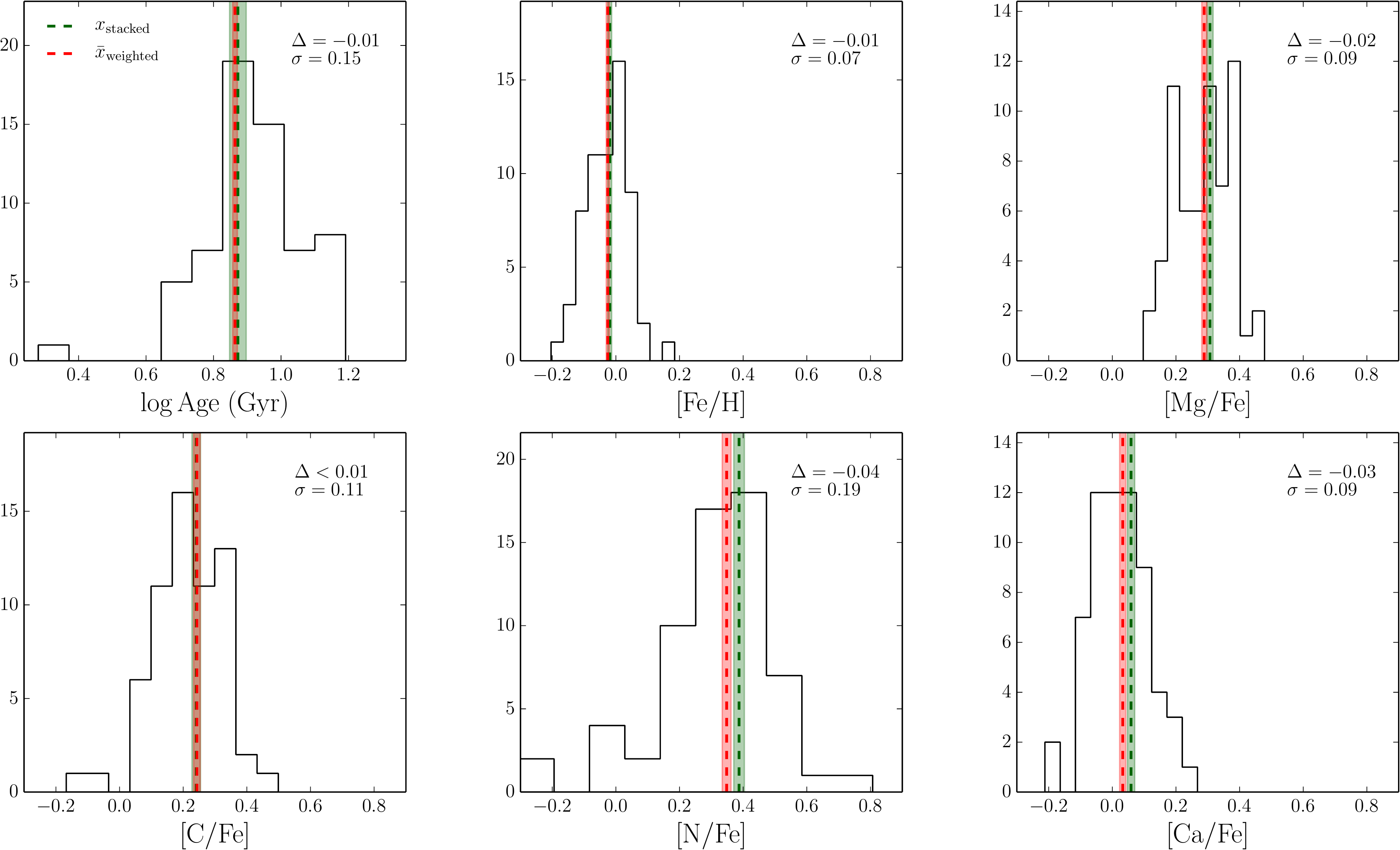}
\caption{Histograms of parameters measured from individual SDSS spectra in the highest mass bin. The green and red lines correspond, respectively, to the best-fit parameter measured from the stacked spectrum and the weighted average of the distribution. Their corresponding uncertainties are shown as shaded regions. The difference between the two quantities and the standard deviation of the distribution are displayed in the top right corner of each panel. The median of each distribution tracks the weighted average very closely, and is thus not displayed. The green and red lines are in excellent agreement, which means that no significant biases are introduced when analyzing stacked spectra.}
\label{fig:stacking_test}
\end{figure*}

In this Appendix we carry out several systematic tests to explore the robustness of our analysis tools and methods. First we explore how our full spectrum fitting model behaves at low S/N. Next we examine the effect of smoothing the spectra to higher effective velocity dispersions. Lastly, we compare the distributions of resulting parameters measured from individual spectra against the results derived from a stacked spectrum. In summary, these systematic tests show that both the age and abundance measurements are generally very robust, resulting in systematic uncertainties of $\lesssim0.05$ dex in most cases.

\subsection{Parameters as a Function of S/N}
\label{section:snr_test}
To test the ability of our models to recover parameters at low S/N, we have artificially degraded high S/N spectra, fit them with our models, and compared the derived fits to the results at high S/N. We select a S/N~$>1000$~\aa$^{-1}$ SDSS stack corresponding to the second highest mass bin to carry out this test. Since the model spectra utilized in the fitting code are constructed from solar-scaled empirical stellar spectral library, we purposefully use the a high-mass bin (second highest rather than the highest due to its sufficiently high S/N) to test whether or not we can successfully recover non-solar abundance measurements. The S/N at 5000~\aa{} in the original spectrum is used as the reference value in the following test. We construct 50 realizations for each S/N value, and compute the difference between the resulting parameter and the best-fit parameter from fitting the high S/N template spectrum. The results are shown in Figure~\ref{fig:snr_test}. The different colored symbols represent two independent error estimates, where the red is the average of the statistical errors for the 50 trials as output by the fitting code, and the blue corresponds to the 1$\sigma$ scatter in the best-fit parameters for the 50 trials. 

There are two remarkable features of this figure. First, the age and [Fe/H] are accurately recovered without significant systematic offsets down to S/N~$\approx10~\text{\aa{}}^{-1}$. [Mg/Fe] and [Ca/Fe], on the other hand, require S/N~$\approx20~\text{\aa{}}^{-1}$, and [C/Fe] and [N/Fe] demand S/N~$\approx30~\text{\aa{}}^{-1}$. Second, the two uncertainty estimates are broadly comparable, though the errors estimated by the spectrum fitting code may be conservative. 

We discuss two caveats regarding the results. First, the results generally become biased toward the mean of the prior at low S/N. In the limit of S/N~$=0~\text{\aa{}}^{-1}$, the fit is unconstrained and the resulting best-fit parameter is trivially the mean of the prior regardless of its true value, with the width of the prior as the uncertainty. Thus the conclusions here are appropriate for the sample in this paper, and should be treated with caution when applied to low S/N spectra of galaxies with unusually extreme stellar population properties. Second, as discussed in Section~\ref{section:keck_results}, we also urge caution with data containing strong systematic uncertainties due to sky subtraction and telluric features. The reliable measurement of stellar population parameters not only hinges on robust models but also on clean rest-frame optical spectra.

Figure~\ref{fig:errorbar_sn} shows the uncertainty on the parameters as a function of S/N per Angstrom. The solid green line is the theoretical expectation $\sigma=(\rm S/N)^{-1}$, and the dashed line labeled J12 is the S/N relation measured in the $g$-band from \cite{Johansson2012}. This relation comes from fitting models to 25 standard Lick absorption line indices measured from SDSS quiescent galaxy spectra. The yellow points are error estimates from fitting a representative sample of individual AGES spectra, and the red symbols are the average of the 50 errors measured by the fitting code (the same red error bars from Figure~\ref{fig:snr_test}). Comparison in this figure between the simulated Gaussian noise result and the real AGES data shows that our assumption of ideal, uncorrelated noise for the former is a reasonable approximation. The derived errors are well-behaved and scale as expected, though the errors are larger probably due to degeneracies. Age and [Fe/H] are reliably measured to $\approx0.1$~dex errors at S/N~$\approx10~\text{\aa{}}^{-1}$, while [Mg/Fe] and [Ca/Fe] are reliably measured to $\approx0.1$~dex errors at S/N~$\approx20~\text{\aa{}}^{-1}$. [C/Fe] and [N/Fe], on the other hand, require S/N~$\approx30\text{--}50~\text{\aa{}}^{-1}$ for $\approx0.1$~dex errors. It is important to note, however, that these errors are statistical only. 

\subsection{Parameters as a Function of Velocity Dispersion}
\label{section:smoothing_test}
Now we turn our attention to exploring systematics arising due to the Doppler broadening of the spectra. To make our test spectra, we divide individual SDSS spectra into two groups ($\sigma < 150$~\kms{} and $\sigma \geq 250$~\kms{}). Spectra in each group are smoothed to $\sigma=150$~\kms{} and $\sigma=350$~\kms{}, continuum-divided, and stacked to create the template spectra. Next, each template spectrum is convolved with a Gaussian kernel whose width is chosen such that $ \sigma_{\rm target}^2 = \sigma_{\rm kernel}^2 + \sigma_{\rm original}^2$. This convolution increases the effective velocity dispersion of the spectrum. The difference between the parameters measured from these smoothed spectra and the original unsmoothed spectrum is shown in Figure~\ref{fig:smoothing_test}. The vertical dotted line marks $\sigma=350$ \kms{}, the effective velocity dispersion to which we broaden our science spectra for the stacking procedure. When $\sigma<150$ \kms{} spectra are smoothed to 350~\kms{}, there is almost no bias for [Mg/Fe], [C/Fe], and [Ca/Fe]. On the other hand, only a modest systematic offset of $\lesssim0.05$~dex is introduced at 350~\kms{} for age, [Fe/H], and [N/Fe]. Moreover, there appears to be relatively small bias introduced in smoothing spectra even up to $\sim1000$~\kms{}, which is quite remarkable. A velocity dispersion of 1000~\kms{} corresponds to a resolution of $R\sim130$. Grism spectra often achieve resolving powers of $\sim100$, suggesting that we may be able to recover reliable abundance patterns even from such low resolution data. 

\subsection{The Effects of Stacking Spectra}
\label{section:stacking_test}
As described in Sections~\ref{section:sdss_sample} and \ref{section:ages_sample}, we bin the individual spectra by mass and redshift, smooth them to an effective velocity dispersion of $\sigma=350$ \kms, and stack the spectra together. The objective is to test whether the stacked spectrum produces results that are consistent with fitting individual, unsmoothed spectra in each bin and then averaging the individual best-fit parameters. To carry out this test, we use the highest-mass SDSS bin for computational feasibility since it contains 62 objects rather than $\sim 3000~\text{--}~20000$ as in other bins. Histograms of parameters measured from fitting individual spectra are shown in Figure~\ref{fig:stacking_test}. The green and red lines correspond, respectively, to the best-fit parameter measured from the stacked spectrum and the weighted average of the distribution. Their corresponding uncertainties are shown as shaded regions. If there were no systematic effects introduced during the stacking process, then we would expect the green and the red lines to overlap to within the uncertainties. The two quantities are in excellent agreement to within 0.05~dex. Weighted average and unweighted average are also comparable to within $\approx0.05$~dex. This test demonstrates that analyzing stacked spectra is essentially equivalent to analyzing individually each spectrum that contributed to the stack.

For the abundances, the width of the distribution is comparable to the measurement uncertainties for the individual galaxies, implying that there is no evidence for an abundance spread within the bin. For the ages, on the other hand, the measurement uncertainties can only account for approximately half of the width of the distribution. Thus there must be an intrinsic variation in the ages of the galaxies within this sample. A more stringent version of the test would involve greater variations within the bin. Exploration of parameter variations within stellar mass bins will be the subject of future work.

\end{appendix}

\newpage
\bibliographystyle{apj}
\bibliography{bibtex.bib}

\newpage
\begin{deluxetable*}{c c c c c c c c c c c}
\tablewidth{0pt}
\tablecaption{Results from Modeling Quiescent Galaxy Spectra \label{table:results}}
\tablehead{\colhead{$z$} & \colhead{$\log M$} & \colhead{$R_{\rm fiber}/R_{\rm e}$} & \colhead{$N_{\rm obj}$} & \colhead{S/N} & \colhead{Age} & \colhead{[Fe/H]} & \colhead{[C/Fe]} & \colhead{[N/Fe]} & \colhead{[Mg/Fe]} & \colhead{[Ca/Fe]}  \\
\colhead{} & \colhead{$\msun{}$} & \colhead{} & \colhead{} & \colhead{\AA{}$^{-1}$} & \colhead{Gyr} & \colhead{} & \colhead{} & \colhead{} & \colhead{} & \colhead{} 
}
\startdata
\sidehead{SDSS}
0.07--0.09 & 10.6 & 1.11 & 19356 & 2457 & 4.38$\;\pm\;$0.01 & $-0.01\;\pm\;$$0.00^{*}$ & 0.18$\;\pm\;0.00^{*}$ & 0.16$\;\pm\;0.00^{*}$ & 0.18$\;\pm\;0.00^{*}$ & 0.00$\;\pm\;0.00^{*}$ \\
\; & 10.9 & 0.76 & 14189 & 2702 & 5.43$\;\pm\;$0.01 & $-0.01\;\pm\;$$0.00^{*}$ & 0.18$\;\pm\;0.00^{*}$ & 0.21$\;\pm\;0.00^{*}$ & 0.21$\;\pm\;0.00^{*}$ & 0.02$\;\pm\;0.00^{*}$ \\
\; & 11.2 & 0.51 & 3059 & 1554 & 6.10$\;\pm\;$0.01 & $-0.02\;\pm\;$$0.00^{*}$ & 0.20$\;\pm\;0.00^{*}$ & 0.30$\;\pm\;0.00^{*}$ & 0.25$\;\pm\;0.00^{*}$ & 0.04$\;\pm\;0.00^{*}$ \\
\; & 11.5 & 0.35 & 62 & 243 & 7.45$\;\pm\;$0.42 & $-0.02\;\pm\;$0.01 & 0.24$\;\pm\;$0.01 & 0.39$\;\pm\;$0.02 & 0.31$\;\pm\;$0.01 & 0.06$\;\pm\;$0.01 \\
\\
\hline
\sidehead{AGES}
0.1--0.2 & 9.9 & 2.97 & 41 & 54 & 2.80$\;\pm\;$0.13 & $-0.05\;\pm\;$0.04 & 0.08$\;\pm\;$0.07 & $-0.38\;\pm\;$0.17 & 0.04$\;\pm\;$0.05 & $-0.06\;\pm\;$0.10 \\
\; & 10.2 & 2.02 & 83 & 111 & 3.57$\;\pm\;$0.17 & $-0.06\;\pm\;$0.02 & 0.16$\;\pm\;$0.03 & $-0.13\;\pm\;$0.07 & 0.13$\;\pm\;$0.03 & $-0.01\;\pm\;$0.04 \\
\; & 10.4 & 1.56 & 192 & 215 & 4.25$\;\pm\;$0.10 & $-0.01\;\pm\;$0.01 & 0.17$\;\pm\;$0.02 & 0.02$\;\pm\;$0.03 & 0.16$\;\pm\;$0.01 & $-0.02\;\pm\;$0.02 \\
\; & 10.7 & 1.06 & 126 & 228 & 5.50$\;\pm\;$0.06 & $-0.03\;\pm\;$0.01 & 0.18$\;\pm\;$0.01 & 0.16$\;\pm\;$0.02 & 0.21$\;\pm\;$0.01 & 0.04$\;\pm\;$0.01 \\
\; & 11.0 & 0.72 & 23 & 119 & 5.84$\;\pm\;$0.14 & 0.02$\;\pm\;$0.01 & 0.16$\;\pm\;$0.03 & 0.18$\;\pm\;$0.04 & 0.23$\;\pm\;$0.02 & 0.01$\;\pm\;$0.02 \\
0.2--0.3 & 10.2 & 2.95 & 34 & 47 & 3.05$\;\pm\;$0.24 & $-0.08\;\pm\;$0.04 & 0.21$\;\pm\;$0.07 & 0.13$\;\pm\;$0.16 & 0.17$\;\pm\;$0.06 & 0.07$\;\pm\;$0.08 \\
\; & 10.5 & 2.0 & 147 & 121 & 3.16$\;\pm\;$0.12 & $-0.06\;\pm\;$0.02 & 0.18$\;\pm\;$0.03 & 0.08$\;\pm\;$0.07 & 0.19$\;\pm\;$0.02 & 0.03$\;\pm\;$0.03 \\
\; & 10.7 & 1.55 & 253 & 196 & 4.24$\;\pm\;$0.11 & $-0.03\;\pm\;$0.01 & 0.14$\;\pm\;$0.02 & 0.21$\;\pm\;$0.03 & 0.20$\;\pm\;$0.01 & 0.01$\;\pm\;$0.02 \\
\; & 11.0 & 1.05 & 88 & 154 & 4.72$\;\pm\;$0.14 & $-0.01\;\pm\;$0.01 & 0.13$\;\pm\;$0.02 & 0.17$\;\pm\;$0.04 & 0.22$\;\pm\;$0.02 & 0.02$\;\pm\;$0.02 \\
\; & 11.3 & 0.71 & 5 & 58 & 6.24$\;\pm\;$0.23 & $-0.05\;\pm\;$0.02 & 0.20$\;\pm\;$0.05 & 0.27$\;\pm\;$0.07 & 0.23$\;\pm\;$0.04 & 0.03$\;\pm\;$0.05 \\
0.3--0.4 & 10.5 & 2.6 & 88 & 72 & 3.27$\;\pm\;$0.21 & $-0.11\;\pm\;$0.03 & 0.19$\;\pm\;$0.05 & 0.26$\;\pm\;$0.10 & 0.18$\;\pm\;$0.04 & 0.04$\;\pm\;$0.06 \\
\; & 10.8 & 1.76 & 260 & 164 & 3.47$\;\pm\;$0.12 & $-0.05\;\pm\;$0.01 & 0.16$\;\pm\;$0.02 & 0.25$\;\pm\;$0.05 & 0.22$\;\pm\;$0.02 & 0.03$\;\pm\;$0.03 \\
\; & 11.0 & 1.36 & 254 & 182 & 4.55$\;\pm\;$0.13 & $-0.02\;\pm\;$0.01 & 0.18$\;\pm\;$0.02 & 0.24$\;\pm\;$0.03 & 0.21$\;\pm\;$0.01 & 0.04$\;\pm\;$0.02 \\
\; & 11.3 & 0.93 & 45 & 85 & 5.61$\;\pm\;$0.15 & $-0.03\;\pm\;$0.02 & 0.19$\;\pm\;$0.04 & 0.26$\;\pm\;$0.05 & 0.29$\;\pm\;$0.03 & 0.04$\;\pm\;$0.04 \\
0.4--0.55 & 10.8 & 2.31 & 144 & 75 & 2.99$\;\pm\;$0.15 & $-0.07\;\pm\;$0.02 & 0.18$\;\pm\;$0.04 & 0.26$\;\pm\;$0.08 & 0.23$\;\pm\;$0.04 & 0.05$\;\pm\;$0.05 \\
\; & 11.1 & 1.57 & 285 & 123 & 3.28$\;\pm\;$0.13 & $-0.04\;\pm\;$0.01 & 0.18$\;\pm\;$0.02 & 0.34$\;\pm\;$0.05 & 0.23$\;\pm\;$0.02 & $-0.03\;\pm\;$0.03 \\
\; & 11.3 & 1.21 & 122 & 89 & 4.00$\;\pm\;$0.18 & $-0.05\;\pm\;$0.02 & 0.14$\;\pm\;$0.03 & 0.25$\;\pm\;$0.06 & 0.30$\;\pm\;$0.03 & 0.08$\;\pm\;$0.03 \\
0.55--0.7 & 10.9 & 2.49 & 24 & 24 & 2.67$\;\pm\;$0.20 & $-0.15\;\pm\;$0.07 & 0.16$\;\pm\;$0.13 & 0.18$\;\pm\;$0.30 & 0.05$\;\pm\;$0.13 & 0.06$\;\pm\;$0.15 \\
\; & 11.0 & 2.18 & 153 & 68 & 2.49$\;\pm\;$0.12 & $-0.02\;\pm\;$0.03 & 0.24$\;\pm\;$0.05 & 0.58$\;\pm\;$0.09 & 0.09$\;\pm\;$0.05 & $-0.03\;\pm\;$0.06 \\
\; & 11.3 & 1.48 & 109 & 67 & 3.06$\;\pm\;$0.11 & $-0.05\;\pm\;$0.03 & 0.26$\;\pm\;$0.04 & 0.73$\;\pm\;$0.09 & 0.19$\;\pm\;$0.04 & 0.00$\;\pm\;$0.05 \\
\\
\hline
\sidehead{Keck}
0.83 & 10.9 & - & 1 & 24 & 5.68$\;\pm\;$0.28 & 0.06$\;\pm\;$0.04 & $-0.32\;\pm\;$0.08 & $-0.54\;\pm\;$0.21 & 0.31$\;\pm\;$0.07 & $-0.23\;\pm\;$0.08 \\
0.83 & 11.1 & - & 1 & 27 & 5.16$\;\pm\;$0.30 & $-0.03\;\pm\;$0.04 & 0.06$\;\pm\;$0.12 & $-0.23\;\pm\;$0.24 & 0.35$\;\pm\;$0.08 & $-0.01\;\pm\;$0.07 \\
\enddata
\tablecomments{Stellar masses are medians within the mass--redshift bins and they assume a \cite{Chabrier2003} IMF. $R_{\rm fiber}/R_{\rm e}$ is estimated using the mass-size relation from \cite{Shen2003} and applying a redshift scaling from \cite{Williams2010} (See Section~\ref{section:gradients}). Signal to noise is the median between 4000~--~5500 \aa{}. Ages are SSP-equivalent ages. The quoted errors are statistical only, and systematic errors are estimated to be $\lesssim0.05$ dex \citep[see the Appendix and also][]{Conroy2014}. Statistical uncertainties smaller than $0.01$ dex are marked with $^{*}$.}
\end{deluxetable*}

\end{document}